
\input lanlmac.tex
\input epsf
\font\cmss=cmss10 \font\cmsss=cmss10 at 7pt
\ifx\epsfbox\UnDeFiNeD\message{(NO epsf.tex, FIGURES WILL BE IGNORED)}
\def\figin#1{\vskip2in}
\else\message{(FIGURES WILL BE INCLUDED)}\def\figin#1{#1}\fi
\def\tfig#1{{\xdef#1{Fig.\thinspace\the\figno}}
Fig.\thinspace\the\figno \global\advance\figno by1}
%


\def\CA {{\cal A}}

\def\CD {{\cal D}}

\def\CF {{\cal F}}

\def\CH {{\cal H}}

\def\CP {{\cal P}}

\def\CT {{\cal T}}

\def\CW {{\cal W}}

\def\CZ {{\cal Z}}
 \def\R{\relax{\rm I\kern-.18em R}}
\font\cmss=cmss10 \font\cmsss=cmss10 at 7pt
\def\Z{\relax\ifmmode\mathchoice
{\hbox{\cmss Z\kern-.4em Z}}{\hbox{\cmss Z\kern-.4em Z}}
{\lower.9pt\hbox{\cmsss Z\kern-.4em Z}}
{\lower1.2pt\hbox{\cmsss Z\kern-.4em Z}}\else{\cmss Z\kern-.4em Z}\fi}
\def\hf{\frac12}

\def\p{\partial}
 
 \def\hf{{1\over 2}}
 \def\11{1\!\! 1}

 \def\pl{{Phys. Lett. }}

\def\np{{Nucl. Phys. }}

\def\a{\alpha}

 \def\l{\lambda}

\def\t{\theta}

\def\k{\eta}
\def\eg{{\it e.g.}}
 \def\ie{{\it i.e.}}

\lref\RABI{S. Elitzur, A. Forge and E. Rabinovici,
\np {\bf B359} (1991) 581. }
\lref\WADIA{G. Mandal, A. Sengupta, and  S. Wadia,
Mod. Phys. Lett. {\bf A6} (1991) 1685.}
\lref\WITTEN{E. Witten, Phys. Rev. {\bf D44} (1991) 314.}
\lref\FZ{Al. B. Zamolodchikov, unpublished;
V.A. Fateev, \pl {\bf B357} (1995) 397.}
\lref\FZZ{V. Fateev, A. Zamolodchikov and  Al. Zamolodchikov,
unpublished.}
\lref\DDK{V. Knizhnik, A. Polyakov and A. Zamolodchikov,  
Mod. Phys. Lett. {\bf A3} (1988) 819;
F. David, Mod. Phys. Lett. {\bf A3} (1988) 1651;
J. Distler and H. Kawai, \np {\bf B321} (1989) 509.}
\lref\KAZMIG{V. Kazakov and A. A. Migdal, \np {\bf B311} (1988) 171.}
\lref\GRKL{D. Gross and I. Klebanov, \np {\bf B344} (1990) 475.}
\lref\GRKLbis{D. Gross and I. Klebanov, \np {\bf B354} (1990) 459.}
\lref\BULKA{D. Boulatov and V.Kazakov, preprint LPTENS 91/24 (1991),
Int. J. Mod. Phys. {\bf A8} (1993) 809, revised version: hep-th/0012228.}
\lref\DASJP{S. Das and A. Jevicki,
Mod. Phys. Lett. {\bf A5} (1990) 1639.}
\lref\DASWD{
S. R. Das, Mod. Phys. Lett. {\bf A8} (1993) 69, 1331;
A. Dhar, G. Mandal and S. Wadia, Mod. Phys. Lett. {\bf A7} (1992) 3703;
{\bf A8} (1993) 1701; A. Dhar \np {\bf B507} (1997) 277.}
\lref\JEV{ A. Jevicki and D. Yoneya, Nucl. Phys. {\bf B411} (1994) 64.}
\lref\MATYTSIN{A. Matytsin and P. Zaugg, hep-th/9611170,
\np {\bf B497} (1997) 658; hep-th/9701148, \np {\bf B497} (1997) 699.}
\lref\MOORE{ G. Moore, hep-th/9203061.}
\lref\On{I. Kostov and M. Staudacher,  \np {\bf B384} (1992) 459.}
\lref\JM{M. Jimbo and T. Miwa, \ Publ. RIMS, Kyoto Univ. {\bf 19}, No. 3
(1983) 943.}
\lref\Hir{R. Hirota,  Direct Method in Soliton Theory, {\it Solitons},
Ed. by R. K. Bullogh and R. J. Caudrey, Springer, 1980.}
\lref\HK{E. Hsu and  D. Kutasov,  hep-th/9212023, Nucl. Phys.
{\bf B396} (1993) 693.}
\lref\UT{K. Ueno and K. Takasaki, Adv. Stud. Pure Math. {\bf 4} (1984) 1;
K. Takasaki, Adv. Stud. Pure Math. {\bf 4} (1984) 139.}
\lref\SAKA{S. Kakei, ``Toda lattice hierarhy and Zamolodchikov's
conjecture", solv-int/9510006 }
\lref\PBM{A. Prudnikov, Yu. Brichkov, O. Marychev, ``Integrals and
Series'', Nauka , Moscow, 1981.}
\lref\DVV{R. Dijkgraaf, E. Verlinde and H. Verlinde,
Nucl. Phys. {\bf B371} (1992) 269.}
\lref\HS{G. Horowitz and A. Strominger, Nucl. Phys. {\bf B360}
(1991) 197; J. Maldacena and A. Strominger, hep-th/9710014.}
\lref\OV{H. Ooguri and C. Vafa, hep-th/9511164, Nucl. Phys.
{\bf B463} (1996) 55.}
\lref\GK{A. Giveon, D. Kutasov and O. Pelc, hep-th/9907178,
JHEP {\bf 9910} (1999) 035;
A. Giveon and D. Kutasov, hep-th/9909110,
JHEP {\bf 9910} (1999) 034; hep-th/9911039,
JHEP {\bf 0001} (2000) 023.}
\lref\TESCH{J. Teschner, hep-th/9712256, Nucl. Phys. {\bf B546}
(1999) 390; hep-th/9712258, Nucl. Phys. {\bf B546} (1999) 369;
hep-th/9906215, Nucl. Phys. {\bf B571} (2000) 555.}
\lref\difkut{P. Di Francesco and D. Kutasov, hep-th/9109005,
Nucl. Phys. {\bf B375} (1992) 119.}
\lref\FATA{ L. D. Fadeev and L. A. Takhtajan, ``Hamiltonian Methods in
the Theory of Solitons'', Springer-Ferlag (1987).}
\lref\SHENKER{S. Shenker, Proceedings of Cargese workshop
on Random Surfaces, Quantum Gravity and Strings, 1990.}
\lref\BRKA{E. Brezin, V. Kazakov and Al. Zamolodchikov,
\np {\bf B338}(1990) 673.}
\lref\PARISI{ G. Parisi, \pl {\bf B238} (1990) 209, 213.}
\lref\GRMI{ D. Gross and N. Miljkovic, \pl {\bf B238} (1990) 217.}
\lref\GIZI{P. Ginsparg and J. Zinn-Justin, \pl {\bf B240} (1990) 333.}
\lref\POLCHINSKI{ J. Polchinski, ``What is String Theory?'',
Lectures presented at the 1994 Les Houches Summer School ``Fluctuating
Geometries in Statistical Mechanics and Field Theory'', hep-th/9411028. }
\lref\GP{G. Gibbons and M. Perry,  hep-th/9204090, Int. J. Mod. Phys.
{\bf D1} (1992) 335; C. R. Nappi and A. Pasquinucci, hep-th/9208002,
Mod. Phys. Lett. {\bf A7} (1992) 3337.}
\lref\YK{I. Kogan, {\it JETP Lett.} 44 (1986) 267, 45 (1987) 709.}
\lref\DOUGLAS{M. R. Douglas,  ``Conformal theory techniques 
in Large $N$ Yang-Mills Theory'', talk at the 1993 Carg\`ese meeting, 
hep-th/9311130.}
\lref\KLEBANOV{I. Klebanov, proceedings of the ICTP 
Spring School on String Theory and Quantum Gravity,
 Trieste, April 1991, hep-th/9108019.}
\lref\KAZREV{V. Kazakov,  ``Bosonic strings and string field theories 
in one-dimensional target space'', proceedings of
Cargese workshop on Random Surfaces, Quantum Gravity and Strings,
1990.}
\lref\ginsmoore{P. Ginsparg and G. Moore, proceedings of TASI
1992, hep-th/9304011.} 
\lref\trtr{ I. Klebanov, {\it Phys. Rev.} {\bf D}51
(1995) 1836. }
\lref\DMP{R. Dijkgraaf, G. Moore, R. Plesser, \np {\bf B394} (1993) 356.}
%
%
\rightline{SPHT-t00/123, LPTENS-00/32, EFI-2000-29}
\Title{
}
{\vbox{\centerline { A Matrix Model }
\centerline{ for the Two Dimensional Black Hole }
 \vskip2pt
}}
%
%
\centerline{Vladimir Kazakov \footnote{$ ^\circ $}{{\tt
kazakov@physique.ens.fr}}}
\centerline{{\it  Laboratoire de Physique Th\'eorique de l'Ecole
Normale Sup\'erieure \footnote{$ ^\ast $}{\rm Unit\'e de Recherche du
Centre National de la Recherche Scientifique et de  l'Ecole Normale
Sup\'erieure et \`a
l'Universit\'e de Paris-Sud.}}}

\centerline{{\ \ \ \it  75231 Paris CEDEX, France}}
\bigskip
\centerline{Ivan K. Kostov 
\footnote{$^\dagger$}{Permanent 
address: Service de Physique Th{\'e}orique, CNRS - URA 2306, 
Commissariat \`a l'\'energie atomique,
C.E.A. - Saclay,  
F-91191 Gif-Sur-Yvette, France.}
\footnote{$^\bullet$}{{\tt kostov@spht.saclay.cea.fr}} 
}
\centerline{\it Yukawa Institute for Theoretical Physics}
\centerline{\it Kyoto University, Kyoto 606-8502, Japan}
 
\bigskip
\centerline{David Kutasov\footnote{$ ^\diamond $}{\tt
kutasov@theory.uchicago.edu
}}
\centerline{\it Department of Physics, University of Chicago,}
\centerline{\it 5640 S. Ellis Avenue, Chicago, IL 60637, USA}
%
 \vskip 1cm
\baselineskip8pt{

\baselineskip12pt{
\noindent
We construct and study a matrix model that describes two
dimensional string theory in the Euclidean black hole
background. A conjecture of V. Fateev, A. and Al. Zamolodchikov,
relating the black hole background to condensation of vortices
(winding modes around Euclidean time) plays an important role
in the construction. We use the matrix model to study quantum
corrections to the thermodynamics of two dimensional black holes.
}}

\Date{December, 2000}

\baselineskip=14pt plus 2pt minus 2pt


\newsec{Introduction }

An interesting solution to the equations of motion of two
dimensional string theory is the semi-infinite cigar 
\refs{\RABI,\WADIA,\WITTEN}, described by the metric and dilaton
\eqn\aaa{\eqalign{
&ds^2={k}\left[dr^2+\tanh^2
r \ d\theta^2\right]\cr
&\Phi-\Phi_0=-2\log\cosh r \cr
}}
where $\theta$ is a periodic coordinate, $\theta\sim\theta+2\pi$,
labeling the location around the cigar, and $r\ge0$ is the
direction along the cigar, with $r=0$ corresponding to the
tip (see fig. 1). The string coupling $e^\Phi$ depends on $r$ -- it
goes to zero far from the tip of the cigar, and attains its
maximal value, $g_s=e^{\Phi_0}$, at the tip. $k$ is a free
parameter which governs the overall size of the cigar.

\bigskip 
\hskip 70pt
\epsfbox{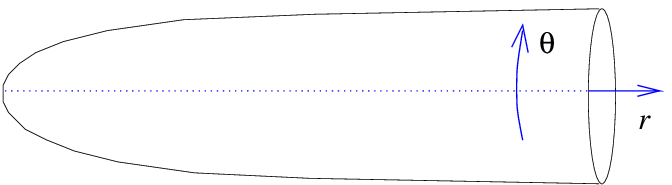}
\vskip 15pt

{\centerline {Fig.1. \ \   The semi-infinite cigar.}}

\bigskip

Some of the contexts in which the geometry \aaa\ appears in string
theory are:

\item{(1)} This geometry can be thought of as a Euclidean version of a 
$1+1$ dimensional black hole \WITTEN. The Minkowski black hole can be 
obtained from \aaa\ by analytically continuing $\theta\to it$.
Furthermore, string propagation in the
geometry \aaa\ can be described by a coset conformal field theory;
the Minkowski version corresponds to the coset $SL(2,R)/U(1)$,
while the Euclidean one is ${SL(2,C)\over SU(2)}/U(1)$. The
parameter $k$ in \aaa\ is the level of the $SL(2)$ current algebra
that enters the construction;\foot{The background \aaa\ is in fact
only valid for large $k$. For finite $k$ there are corrections
that were found in \DVV.} the mass of the black hole, $M$, is
encoded in the value of the string coupling at the tip of the cigar,
$M\propto \exp(-2\Phi_0)$.

\lref\kkss{D. Kutasov and D. A. Sahakyan, hep-th/0012258.}

\item{(2)} The near-horizon geometry of non-extremal $NS5$-branes
is a tensor product of the two dimensional black hole geometry \aaa,
$R^5$ (the spatial directions along the fivebranes) and $S^3$ (the
angular directions transverse to the fivebranes) \HS. The
corresponding
worldsheet theory is a supersymmetric version of the above coset
CFT. In this case $k$ is the number of $NS5$-branes, and the string
coupling at the tip of the cigar measures the energy density of the
fivebranes (see \kkss\ for a recent discussion).

\item{(3)} String theory on a Calabi-Yau manifold near a point
in moduli space at which the manifold develops an isolated
singularity is described by a geometry that includes \aaa\ as
a factor \refs{\OV,\GK}. The level $k$ depends in this case
on the particular singularity, and the string coupling at the
tip of the cigar corresponds to a non-perturbative energy scale
in the theory -- the energy of D-branes wrapped around cycles
whose size vanishes at the singularity.

\noindent
Further study of the geometry \aaa\ would thus be useful for
understanding the physics associated with horizons and singularities
in string theory. In particular, one would like to obtain a
better understanding of the thermodynamics and statistical
mechanics of the corresponding black holes (or black branes).

A promising framework in this regard is holography. In all three 
examples mentioned above, string theory in the geometry \aaa\ is 
believed to be holographically dual to a non-gravitational theory.  
In the case (1) the dual is the ``old matrix model'' -- Matrix 
Quantum Mechanics (MQM) in the double scaling limit -- while in 
the other two examples it is Little String Theory (LST).

In this paper we will focus on the first example, the $1+1$
dimensional black hole.  We will propose a solvable formulation 
of two dimensional string theory in the background \aaa, based on 
MQM.  This approach to two dimensional string theory in flat 
spacetime with a condensate of the tachyon field \KAZMIG\ led 
to a solution of the theory to all orders in string perturbation 
theory \refs{\BRKA,\PARISI,\GRMI,\GIZI}; see 
\refs{\KAZREV,\KLEBANOV,\ginsmoore}\ for reviews. The collective field
theory of the singlet sector of MQM was identified with the
dynamics of the tachyon field in this background \DASJP.  Some
early attempts to find a matrix realization (in the singlet
sector of the matrix model) of two dimensional string theory in
the background \aaa\ are described in \refs{\DASWD,\POLCHINSKI,\JEV}.

As mentioned above, one of the main motivations for looking
for a holographic description is to study the statistical mechanics
of these black holes. Standard black hole thermodynamics
arguments imply that there is a Hagedorn density of black 
holes:
\eqn\rhoeb{\rho(M)\sim M^{s_1} e^{\beta_H M}.}
The inverse temperature $\beta_H$ can be read off the geometry 
\aaa, while $s_1$ is given by a one loop calculation in string 
theory. The prediction \rhoeb\ is particularly interesting since, 
at least perturbatively, two dimensional string theory has very few 
states (one field theoretic degree of freedom in $1+1$ dimensions). 
Thus, it is interesting to verify the thermodynamic prediction \rhoeb\ 
by a direct counting of states. In other examples, this proved easier 
to do in the holographically dual picture, and one might hope that the 
same will happen here.

The starting point of our analysis is an observation made by
V. Fateev, A. Zamolodchikov and Al. Zamolodchikov \FZZ. These
authors conjectured  that the $SL(2)/U(1)$ coset CFT is equivalent
to the Sine-Liouville model, \ie\ $c=1$ CFT coupled to a Liouville
field, with the cosmological constant tuned to zero and the scale
set by a winding mode of the $c=1$ field. This correspondence has
not been proven but evidence for its validity has been
presented\foot{The Sine-Liouville model is the massless limit of the
``sausage model'' considered in \FZ.} \FZZ. Our main purpose
here will be to define and study the Sine-Liouville background
of two dimensional string theory using MQM\foot{A direct argument in favour of the 
description of the 2d black hole in terms of compact MQM, based on 
the similar conjecture relating thermal YM to the ADS black
hole,  was given to us by A. Polyakov in 1998.} .

{}From the point of view of the usual Liouville background
of two dimensional string theory, the Sine-Liouville vacuum
corresponds to condensation of vortices (winding modes).
We will treat the vortices as a perturbation, and will show
that in the cases of interest, the process of taking the strength
of the perturbation to infinity (corresponding to vortex
condensation) can be
studied using MQM. Some steps in this direction were taken in the
past. In particular, G. Moore \MOORE\ studied the Sine-Gordon model
coupled to gravity by treating the Sine-Gordon term as a perturbation.
The resulting structure was further discussed in \HK\ where it was
shown that the model describes a renormalization group flow from the
$c=1$ model coupled to gravity in the UV, to a set of decoupled
$c=0$ models coupled to gravity in the IR. As we will see below,
in a certain region of the parameter space of the model, there
is another possible critical point -- the Sine-Liouville model.

In the matrix model formulation, it was conjectured
\refs{\GRKL,\GRKLbis} and then demonstrated \BULKA\
that vortices on the worldsheet correspond to $U(N)$
non-singlet states in MQM. Their wave functions are
described by Young tableaux
of $U(N)$; the vortex charge is equal to the number of boxes.
It is also possible to study vortices, at least in the spherical
approximation, using the dual matrix description based on
discrete time \refs{\PARISI,\GRKLbis,\MATYTSIN}.

By using the connection between vortices and non-singlet states
in MQM, we will construct an integrable
system, the infinite Toda chain hierarchy, that interpolates
between the usual $c=1$ string and Sine-Liouville (or two
dimensional black hole \FZZ) backgrounds 
 structure allows one to compute the partition sum and correlation
functions of the theory to all orders in string perturbation
theory. As an example we will give explicit expressions for the
partition functions on the sphere and on the torus\foot{Toda integrable structure 
of the   $c=1$ string theory perturbed by purely tachyon source
has been first discovered by Dijgraaf, Moore and Plesser in
 \DMP. Our description is related to their by T-duality.}.

The plan of the paper is as follows. In section 2 we review the
precise statement of the correspondence between the $SL(2)/U(1)$ coset
CFT and the Sine-Liouville model, and some of the evidence for it. In
section 3 we use the results of \refs{\MOORE,\HK} to study the $c=1$
string theory (at tree level) in the presence of the
Sine-Liouville interaction.  We show that in a certain range of the
parameters defining the model, there is an obstruction to turning off
the cosmological constant and using the Sine-Liouville interaction to
set the scale (associated with an instability of Sine-Liouville theory),
but in the range of parameters relevant for the correspondence of
section 2 this obstruction disappears and the Sine-Liouville/black
hole phase is stable.

In section 4 we introduce the MQM describing vortices and reduce
it to a one-matrix model whose degrees of freedom correspond to
the matrix holonomy factor around the compactification circle.
We find that the partition function of this model is a
$\tau$-function of the infinite Toda chain hierarchy\foot{The fact that the 
generating function for the tachyon amplitudes in the $c=1$ string theory
posesses  Toda lattice symmetry has been established in \DMP.} 
 We prove this
statement for a more general partition function, which can be thought
of as the generating function of all (multi-)vortex correlators in
two dimensional string theory.

In Section 5 we use the Toda chain equations and the KPZ-DDK scaling
properties of the partition function to obtain ordinary
differential equations for the genus $h$ partition functions.
We solve these equations explicitly for the sphere ($h=0$), thus
proving a conjecture of \MOORE, and for the torus ($h=1$) which is a
new result.

In section 6 we discuss the implications of our results 
for black hole thermodynamics. In particular we compute
the coefficient $a$ in \rhoeb\ and show that the Hagedorn
temperature is associated with a phase transition. In section 7 we
briefly describe the thermodynamics of black holes in fermionic two
dimensional string theory, which helps to clarify some aspects of the
bosonic analysis. In section 8 we comment on our results and describe
some open problems. Some of the technical details are contained in the
appendices.

\newsec{On the equivalence of the cigar and Sine-Liouville CFT's}

As mentioned above, the Euclidean black hole worldsheet
CFT \aaa\ corresponds to the coset
\eqn\cosetconf{H_3^+/U(1);\;\;H_3^+\equiv{SL(2,C)\over SU(2)}.}
It can be studied using standard coset CFT techniques, and is
closely related to CFT on Euclidean $AdS_3(=H_3^+)$; see \eg\
\TESCH. Below we list some properties of this theory.

The central charge of \cosetconf\ is given in terms of the level
of the underlying $SL(2)$ current algebra, $k$, by
\eqn\ccos{c={3k\over k-2}-1.}
In particular, for $k=9/4$, $c=26$, and the coset is a good
classical solution of two dimensional bosonic string theory.
An important set of observables corresponds to momentum and
winding modes on the cigar, $V_{j;m,\bar m}$. These operators
have scaling dimensions
\eqn\scdimv{\Delta_{j;m,\bar m}=-{j(j+1)\over k-2}+{m^2\over k};\;\;
\bar\Delta_{j;m,\bar m}=-{j(j+1)\over k-2}+{\bar m^2\over k},}
where $m,\bar m$ are given by
\eqn\mbarm{m={1\over2}(n_1+n_2k);\;\;\bar m=-{1\over2}(n_1-n_2k);
\;\;n_1,n_2\in Z.}
One can think of the integers $n_1$ and $n_2$ as momentum and winding 
around the cigar (in the $\theta$ direction in figure 1). As is clear 
from the geometry of the cigar, the momentum, $m-\bar m$, is conserved 
in this CFT, while the winding, $m+\bar m$, is not.

Spherical two and three point functions of the observables $V_{j;m,\bar m}$
were computed in \refs{\FZZ,\TESCH}. The two point function is
(suppressing the dependence on the separation of the worldsheet
locations of the vertex operators)
\eqn\twoptvjm{\langle V_{j;m,\bar m}V_{j;-m,-\bar m}\rangle=(k-2)
[\nu(k)]^{2j+1}{\Gamma(1-{2j+1\over k-2})\Gamma(-2j-1)\Gamma(j-m+1)
\Gamma(1+j+\bar m)\over\Gamma({2j+1\over k-2})\Gamma(2j+2)
\Gamma(-j-m)\Gamma(\bar m-j)},}
where
\eqn\nuk{\nu(k)\equiv{1\over\pi} {\Gamma(1+{1\over k-2})\over
\Gamma(1-{1\over k-2})}.}
The result for the three point function is more involved
and can be found in \TESCH.

String perturbation theory is an expansion in $\exp(-2\Phi)$.
On the cigar, the dilaton $\Phi$ varies as in \aaa, but string
perturbation theory still makes sense -- it can be thought of as
an expansion in $\exp(2\Phi_0)$, the value of the dilaton at the tip
(which is proportional to $1/M$). Thus, the genus expansion on
the cigar is a $1/M$ expansion. 

The dual Sine-Liouville theory \FZZ\ is described by the
Lagrangian (we set $\alpha'=1$)
\eqn\sineliouv{L={1\over4\pi}\left[(\partial x)^2 +(\partial\phi)^2
+Q\hat R\phi+\lambda e^{b\phi}\cos R(x_L-x_R)\right].}
The central charge of this theory is $c=1+6Q^2+1$; comparing
to \ccos\ we find that
\eqn\formq{Q^2={1\over k-2}.}
The radius of $x$ is $R=\sqrt k$, the same as the radius
of $\theta$ far from the tip of the cigar in \aaa. The cosine
interaction in \sineliouv\ is thus the lowest lying winding mode,
with winding number equal to one. Requiring that the scaling dimension
of the sine-Liouville interaction is equal to one, leads to
\eqn\dimone{{1\over4} R^2-{1\over4}b(b+2Q)=1}
with the solution
\eqn\bbb{b=-{1\over Q}=-\sqrt{k-2}.}
Very far from the tip of the cigar, \aaa\ describes a cylinder
with a dilaton field that varies linearly along it. The same is
true for the geometry corresponding to \sineliouv\ far from the
potential wall (\ie\ at $\phi\to\infty$). The two cylinders can
thus be identified (for large $k$) via
\eqn\identpar{\eqalign{r&\sim - Q\phi\cr
                       \theta&\sim {x\over \sqrt{k}}.\cr }}
The non-trivial statement of \FZZ\ is that the two CFT's
also agree when one includes the interaction that cuts
off the large string coupling region $\phi\to-\infty$. This
is far from obvious, since in one case the strong coupling
region is eliminated by changing the topology of the cylinder
to that of the cigar \aaa, while in the other this is achieved
by turning on the potential in \sineliouv.

The relation between the coset and Sine-Liouville CFT's is a
strong-weak coupling duality on the worldhseet. The cigar CFT
becomes weakly coupled in the limit $k\to\infty$, where the
geometry \aaa\ becomes weakly curved. In that region the
wavefunction of the Sine-Liouville potential, which has the
large $\phi$ behavior
\eqn\wavesl{\Psi(\phi)\sim e^{(Q-{1\over Q})\phi}}
goes rapidly to zero as $\phi\to\infty$, \ie\ $\Psi$ is
supported in the region of small $\phi$ and the theory
is strongly coupled. The semiclassical limit in Sine-Liouville
is $Q\to\infty$ (\ie\ $k\to 2$ \formq). In this case the
wavefunction \wavesl\ is supported at $\phi\to\infty$, far
from the potential wall \sineliouv\ and the theory is weakly
coupled. The cigar is highly curved in this limit; thus the
coset CFT is strongly coupled.

The operators $V_{j;m,\bar m}$ mentioned above are described
in the Sine-Liouville model as\foot{More precisely, this is
the form of the vertex operators as $\phi\to\infty$.}
\eqn\mapops{V_{j;m,\bar m}\leftrightarrow
e^{ip_Lx_L+ip_Rx_R+\beta\phi},}
where
\eqn\plr{\eqalign{
&p_L={n_1\over R}+n_2R\cr
&p_R={n_1\over R}-n_2R\cr
&\beta=2Qj.\cr
}}
Indeed, the scaling dimensions
\eqn\deltalr{
\Delta={1\over4}p_L^2-{1\over4}\beta(\beta+2Q);\;
\bar\Delta={1\over4}p_R^2-{1\over4}\beta(\beta+2Q)}
agree with \scdimv. Eq. \plr\ also makes it clear
that $n_1$ and $n_2$ are the momentum and winding
around the cylinder (or cigar). Note also that like
in the cigar CFT, in \sineliouv\ the momentum $p_L+p_R$
is conserved, while the winding number $p_L-p_R$ is
broken. In the cigar CFT the reason for that is
that winding can slip off the tip of the cigar;
in \sineliouv\ the interaction breaks this symmetry
explicitly.

The Sine-Liouville model \sineliouv\ has similar scaling
properties to Liouville theory. For example, the standard
scaling analysis \DDK\ implies that the partition sum has
the following perturbative expansion:
\eqn\pertz{\CF(\lambda, g_s)=\sum_{h=0}^\infty \CF_h
\left(g_s\lambda^{-{1\over k-2}}\right)^{2(h-1)},}
where $\CF_h$ is the genus $h$ partition sum, and
$g_s$ is the string coupling constant.\foot{defined as
the value of $\exp\Phi$ at some arbitrary point along
the cylinder.} As in Liouville theory, the physics depends
only on the combination $g_s\lambda^{-{1\over k-2}}$, which
can be thought of as the effective string coupling in the
background \sineliouv. From now on, we will absorb
$g_s$ into $\lambda$ (\ie\ set $g_s=1$).

We see that the string coupling expansion in Sine-Liouville
theory is a large $\lambda$ expansion, like in ordinary
Liouville theory. In string theory on the cigar, as mentioned
above, the string coupling at the tip of the cigar is related
to the mass of the black hole via $g_s^2\sim 1/M$ where $M$ is
measured in Planck units. Therefore, in the correspondence of
\FZZ, $\lambda$ is related to the mass of the black hole, $M$,
\eqn\mapcoup{M \leftrightarrow \lambda^{2\over k-2}.}
An important piece of evidence for the equivalence of the cigar
and Sine-Liouville CFT's is the agreement of a large class of
two and three point functions on the sphere in the two models
\FZZ. We will next review this agreement for the two point function;
the three point functions can be discussed along the same lines.

The two point functions of $V_{j;m,\bar m}$ in the coset CFT,
given by \twoptvjm, exhibit a series of poles. As explained
in \GK\ (for the supersymmetric case; a similar analysis
applies to the bosonic case discussed here), the poles of
the first $\Gamma$ function in the numerator play a different
role from those of the rest of the factors. These poles occur
at values of $j$ for which
\eqn\polesj{1-{2j+1\over k-2}=-n;\;\;n=0,1,2,\cdots.}
Near such a pole, the amplitude has the form

\eqn\resvjm{\langle V_{j;m,\bar m}V_{j;-m,-\bar m}\rangle\sim{(k-2)
[\nu(k)]^{2j+1}(-1)^n\over n!(n+1-{2j+1\over k-2})}
{\Gamma(-2j-1)\Gamma(j-m+1)\Gamma(1+j+\bar m)\over
\Gamma({2j+1\over k-2})\Gamma(2j+2)\Gamma(-j-m)\Gamma(\bar m-j)}
.}
As observed in \FZZ, the poles \polesj\ are also obtained in
Sine-Liouville theory \sineliouv, and the residues agree
with \resvjm. We next review this calculation.

The two point function \twoptvjm\  corresponds in Sine-Liouville to
\eqn\twoptsl{\CA(p_L,p_R)\equiv\langle e^{ip_Lx_L+ip_Rx_R+\beta\phi}
e^{-ip_Lx_L-ip_Rx_R+\beta\phi}\rangle}
(see \mapops, \plr).
The theory \sineliouv\ is interacting, and in general
such correlation functions are difficult to compute
(although much progress has been made on this problem
in the last ten years). However, in some cases one can
compute the residues of certain poles (which will turn
out to correspond precisely to \polesj) as follows
(see \eg\ \difkut\ for a review). Split the path integral
over the Liouville field $\phi$ into a zero mode integral
and a path integral over the non-zero modes. Performing the
integral over the zero mode, one formally finds (after
absorbing an overall constant into the definition of the
path integral)
\eqn\bulkam{\CA(p_L,p_R)=\lambda^s\Gamma(-s)
\langle e^{ip_Lx_L+ip_Rx_R+\beta\phi}
e^{-ip_Lx_L-ip_Rx_R+\beta\phi}
\left[\int e^{b\phi}\cos R(x_L-x_R)\right]^s\rangle_{\lambda=0}
.}
The expectation value in \bulkam\ is understood to
exclude the zero mode of $\phi$; note also that it
is performed with the free action ($\lambda=0$ in
\sineliouv). $s$ is the KPZ scaling parameter, which
using \bbb, \plr\ is
\eqn\sss{s={2(2j+1)\over k-2}.}
In general the representation
\bulkam\ is highly formal since it is difficult to
make sense of the non-integer power of the
interaction in the correlator.
In situations where $s$ is a non-negative integer, the
correlator in \bulkam\ does seem to make sense but the prefactor
diverges. The nature of the divergence is well understood \difkut.
Amplitudes with $s\in Z_+$ are {\it bulk} amplitudes -- they
correspond to processes that can occur anywhere in the infinite
region far from the Sine-Liouville wall and the divergence
in question is nothing but the volume of that region.\foot{Note that
these amplitudes are in general sensitive to the structure of 
the wall, since they involve bulk interactions between the incoming 
quanta and those that make up the wall. Amplitudes with $s=0$ are
insensitive to the structure of the wall, a fact that will play a role
later.}
Regularizing the theory by cutting off the region $\phi\to\infty$
(far from the wall \sineliouv) amounts to replacing
\eqn\regbulk{\lambda^s\Gamma(-s)\to {(-1)^{s+1}\over
s!}\lambda^s\log{\lambda\over\Lambda},}
where $\Lambda$ is a UV cutoff.

To summarize, the residues of the poles of $\Gamma(-s)$ in \bulkam\
are given by bulk amplitudes, which have an integral representation
of Shapiro-Virasoro type. These can be computed by standard
techniques.

When $s$ is odd, the bulk amplitude vanishes. The
reason is that winding number, which is broken by the Sine-Liouville
interaction, is actually a good symmetry in the bulk of space, and
for $s\in 2Z+1$ the correlator \bulkam\ does not conserve winding.
Thus, only poles with
\eqn\polessl{s=2(n+1);\;\;n=0,1,2,\cdots}
survive (the amplitude with $s=0$ is special and can be shown to
vanish in this case).
Comparing to \polesj\ we see that the amplitude $\CA(p_L,p_R)$ 
\bulkam\ has poles at the same places as \twoptvjm\ (the rest of 
the poles of \twoptvjm\ do not correspond to bulk physics; they 
have to do with the existence of certain normalizable states living 
near the tip of the cigar).

Near the poles, \bulkam\ is given by
\eqn\ressl{\eqalign{&\CA(p_L,p_R)\sim
{\lambda^{2(n+1)}\over 2^{n+1}(2n+2)!(n+1)!^2
(n+1-{2j+1\over k-2})}\cr
&\langle e^{ip_Lx_L+ip_Rx_R+\beta\phi}
e^{-ip_Lx_L-ip_Rx_R+\beta\phi}
\left[\int e^{b\phi}e^{i R(x_L-x_R)}\right]^{n+1}
\left[\int e^{b\phi}e^{-i R(x_L-x_R)}\right]^{n+1}
\rangle.\cr}}
As usual, all but three of the vertex operators in \ressl\
are integrated over the complex plane. These integrals can
be evaluated (they are generalizations of those discussed
\eg\ in \difkut), and shown to agree with \resvjm. We will
not discuss the details of these calculations here.

One might worry that this agreement is coincidental since
one can rescale the operators $V_{j;m,\bar m}$ by a
$(j;m,\bar m)$ dependent factor. However, the relative normalization
of the operators entering the correspondence \mapops\ is
in fact fixed by requiring that their wave-functions coincide
in the region very far from the tip of the cigar or Sine-Liouville
potential (\eg\ they can be taken to be incoming waves of unit strength).
This makes the comparison of the two-point functions, which
upon analytic continuation to Minkowski spacetime correspond
to the scattering matrices off the tip of the cigar or Sine-Liouville
potential, meaningful. The authors of \FZZ\ also showed that
the three point functions of the operators \mapops\ exhibit similar
poles, and the residues again agree.

\lref\kutsei{D. Kutasov and N. Seiberg, Phys. Lett. {\bf B251} (1990)
67; for a review, see D. Kutasov, hep-th/9110041.}
\lref\kazsuz{Y. Kazama and H. Suzuki, Nucl. Phys. {\bf B321}
(1989) 232.}

It is useful to mention for future reference that the correspondence
between the cigar and Sine-Liouville conformal field theories has a
(worldsheet) supersymmetric generalization proposed in \GK. It
relates the $N=1$ superconformal  coset model $SL(2)/U(1)$ to the
$N=2$ Liouville theory\foot{It might seem surprising
that an $N=1$ superconformal field theory can be equivalent to an
$N=2$ superconformal one. In fact, it turns out that the $N=1$
supersymmetric $SL(2)/U(1)$ coset has an accidental $N=2$
supersymmetry. It is a special case of the Kazama-Suzuki construction
\kazsuz.} (see \eg\ \kutsei\ for a description of $N=2$ Liouville).

\newsec{ Sine-Liouville as a perturbation of the
$c=1$ string}

As discussed in section 2, one approach to studying the
black hole background in two dimensional string theory
is to study the dual Sine-Liouville theory \sineliouv.
This is what we will do below.

In bosonic two dimensional string
theory the central charge \ccos\ is $c=26$, which determines
the different parameters that appeared in section 2 to be
$k=9/4$, $Q=2$, $R=3/2$. We would like to use the matrix
model approach to analyze this model.
An immediate problem that one runs into is that in the matrix
model $\lambda$ is set to zero, and instead the Lagrangian
contains the Liouville potential $\mu\phi\exp(-2\phi)$.
One can attempt to define \sineliouv\ by means of conformal
perturbation theory, \ie\ consider the theory with Lagrangian
\eqn\conepert{L={1\over4\pi}\left[(\partial x)^2 +(\partial\phi)^2
+2\hat R\phi+\mu\phi e^{-2\phi}+\lambda e^{(R-2)\phi}
\cos R(x_L-x_R)\right]}
and treat $\lambda$ as a perturbation. General arguments lead
one to expect that the radius of convergence of the resulting
series is finite. Since we are interested in a continuation
to $\lambda\to\infty$ (or equivalently $\mu\to 0$), we are
not guaranteed that the series will converge. Indeed, we will
see that in some cases there is an obstruction to taking this
limit -- one encounters a singularity at finite (real) $\lambda$.

This is not unexpected; the classical Lagrangian \conepert\
with $\mu=0$ has a potential which is not bounded from below.
Thus, at least for small $R$, where the cosine is slowly varying
and the semiclassical picture is reliable, we expect the limit
$\mu\to 0$ not to exist. As $R$ increases, quantum effects (on the
worldsheet) become important, and the vacuum energy could in principle 
be renormalized in such a way that the limit exists. We will
see below that this is indeed what happens. Fortunately,
for $R=3/2$, the value relevant for the correspondence with the
cigar, there is no singularity and the Sine-Liouville model
exists. Of course, this has to be the case if the conjecture
described in section 2 is correct.

Consider, for example, the partition sum of \conepert\ on the
sphere. Define
\eqn\tachp{\CT_p\equiv{\Gamma(p)\over\Gamma(1-p)}
\int e^{ip(x_L-x_R)+(|p|-2)\phi}.}
After rescaling $\lambda$ appropriately, the partition sum
corresponding to \conepert\ becomes
\eqn\partm{F_0(\lambda_\pm,\mu)=\langle e^{-\lambda_+\CT_R-\lambda_-
\CT_{-R}}\rangle.}
We are interested in the case
$\lambda_+=\lambda_-={1\over2}\lambda$,
but it will be convenient to keep $\lambda_\pm$ independent.
Expanding in $\lambda_\pm$ and using winding number conservation,
we find
\eqn\lamper{F_0(\lambda,\mu)=\sum_{n=0}^\infty{(\lambda_+
\lambda_-)^n\over n!^2}\langle \CT_R^n\CT_{-R}^n\rangle_0.}
The coefficients in this series are $2n$ point functions
on the sphere, which can in principle be computed in the
matrix model. G. Moore \MOORE\ conjectured a general form
for these amplitudes based on an extrapolation of matrix
model results from small $n$:
\eqn\MOOR{\langle \CT_{R}^{n} \CT_{-R}^{n} \rangle_0 =
 R\  n! \ \mu^{2 } [(1-R) \mu^{R-2}]^{n}
{\Gamma(n(2-R)-2)\over  \ \Gamma(n(1-R)+1)};\;(n\ge 1).}

The term with $n=0$ is the usual $c=1$ partition sum on the sphere,
\eqn\zzero{F_0(\lambda=0,\mu)=-{R\over2}\mu^2\log{\mu\over\Lambda}.}
Combining it with \MOOR\ we find
\eqn\zolmu{F_0(\lambda,\mu)=-{R\over2}\mu^2\log{\mu\over\Lambda}+R\mu^2
\sum_{n=1}^\infty{1\over n!}\left[(1-R)\mu^{R-2}\lambda_+
\lambda_-\right]^n
{\Gamma(n(2-R)-2)\over \ \Gamma(n(1-R)+1)}.}
Differentiating twice with respect to $\mu$ (and
dropping a non-universal constant) we find
\eqn\suscept{\chi_0(\lambda,\mu)\equiv\partial_\mu^2
F_0(\lambda,\mu)= -R\log\mu+R
\sum_{n=1}^\infty{(-z)^n\over n!}
{\Gamma(n(2-R))\over \ \Gamma(n(1-R)+1)},}
where
\eqn\defzz{z\equiv(R-1)\lambda_+\lambda_-\mu^{R-2}}
measures the ``dimensionless strength'' of the perturbation
$\lambda$. The series in \suscept\ can be summed using
equation 5.2.13.30 in \PBM:
\eqn\ppbbmm{\sum_{n=0}^{\infty} {\Gamma(n a +b -1)\over n!
\Gamma(n(a-1)+b)} (-z)^n
= {(1-s)^{b-1} \over b-1},
\qquad  {s\over (1-s)^a}\equiv z.}
Taking the limit $b\to 1$ and plugging in \suscept\
(with $a=2-R$) leads to
\eqn\finsuscep{\chi_0(\lambda,\mu)=-R\log\mu+R\log (1-s),}
where $s$ is given by
\eqn\xxyy{{s\over (1-s)^{2-R}}=z.}
For small $\lambda$, expanding \xxyy\ one finds
$s\simeq z+O(z^2)$. The large $z$ behavior depends on
$R$. Next we analyze this behavior as a function of $R$.

We restrict attention to $R\le 2$. The reason is
that for $R>2$ the $\lambda$ perturbation in \conepert\
qualitatively changes the behavior of the potential in the
weak coupling (UV) region $\phi\to\infty$. Instead of dying
off, it now fluctuates with an amplitude that diverges as
$\phi\to\infty$. This is a reflection of the fact that in
flat space (\ie\ discarding $\phi$), the $\lambda$ perturbation
is non-renormalizable in this case.

It turns out that there is also a difference between the
behavior for $R<1$ and $R>1$. Consider first the case $R<1$.
The parameter $z$ \defzz\ is negative in this case and we
would like to analyze the behavior as $z\to-\infty$. However,
looking at \xxyy\ we see that there is an obstruction:
$\partial_s z$ is positive only for $s>s_c=-1/(1-R)$. Thus,
varying $s$ along the negative real axis we can not probe 
$z<z_c= -(1-R)^{1-R}/(2-R)^{2-R}$.

The physics of this obstruction was explained in \HK.
It corresponds to $c=0$ critical behavior which is
expected to occur in the model \conepert. The $c=1$ field
$x$ (more precisely the T-dual field $\tilde x=x_L-x_R$)
settles into one of the minima of the cosine potential
and decouples, leaving behind a $c=0$ system (coupled
to gravity). While this
is more clearly visible in the fixed area representation
(see \HK\ for the details), it can also be seen directly
from \finsuscep, \xxyy. Near the critical point we have
$z-z_c\sim (s-s_c)^2$. Thus the leading singular behavior
of \finsuscep\ as $z\to z_c$ is $\chi_0\sim (z-z_c)^{1\over2}$,
which is the expected behavior for pure gravity.

Thus it appears that for $R<1$ the Sine-Liouville theory \sineliouv\
is unstable (as expected on general grounds -- see the discussion
after eq. \conepert) and can not be reached\foot{We note for completeness
that we could discuss the system with {\it positive} $z$, in which
case there is no obstruction to taking $z\to\infty$. 
As is clear from \xxyy, as $z\to\infty$,
$s\to 1$, and we seem to find a new fixed point. Positive $z$
corresponds for $R<1$ to imaginary $\lambda$ (see \defzz). Thus, for
$R<1$ one {\it can} study an analog of Sine-Liouville, but
with an imaginary
potential. This theory is non-unitary. Also, for $R<1$ the winding
number two perturbation is relevant and one can study flows involving it.}  
from \conepert\ by taking $\mu\to 0$.

We next turn to $1<R<2$. In \HK\ it was shown that quantum
corrections generate in this case a finite positive contribution
to the effective cosmological constant. Therefore, one expects
the limit $\mu\to 0$ to be non-singular. This is indeed visible
in the formulae above. In this case $z>0$ \defzz\ and the relation
between $z$ and $s$ \xxyy\ is monotonous. Thus, as $z\to\infty$,
$s\to 1$. This is the critical point we are interested in, in
which $\mu$ has been effectively switched off. As a check, note
that in this limit the dependence of the correlator $\chi_0$
on $\mu$ drops out, as it should. Indeed, 
combining \finsuscep\ and \xxyy\ and using \defzz\ we 
find the following algebraic equation for the 
susceptibility:
\eqn\susca{\mu e^{{\chi_0\over R}} +(R-1) \lambda_+\lambda_-
 e^{ {2-R\over R} \chi_0}=1.}
{}From here we find that at large $\lambda$
\eqn\ccchhh{\chi_0\simeq{R\over R-2}\log\left[(R-1)\lambda_+\lambda_-
\right].}
This is easily seen to be the correct KPZ scaling for the two point
function of $e^{-2\phi}$ in Sine-Liouville theory.

One can in fact go further and find the whole small $\mu$ (or large
$\lambda$) expansion of the susceptibility.\foot{To do this it is 
useful to note that \susca\ is invariant under: 
$2-R \leftrightarrow {1\over 2-R}$, $\sqrt{2-R}\log\mu
\leftrightarrow {1\over\sqrt{2-R}}\log[(R-1)\lambda_+\lambda_-]$, 
${\sqrt{2-R}\over R}\chi_0\leftrightarrow{1\over R\sqrt{2-R}}\chi_0$.}
Namely, from \finsuscep\ and \xxyy\ we obtain the following 
parametrization of the susceptibility
\eqn\finsuscept{\chi_0(\lambda,\mu)=-R\log\mu+R\log t=-
{R\over 2-R} \log\left((R-1)\lambda_+\lambda_-\right)+ {R\over
2-R}\log(1-t),} where $t(=1-s)$ is given by
\eqn\ssyy{{t\over (1-t)^{1\over 2-R}}=y}
and 
\eqn\defyy{y\equiv z^{-{1\over 2-R}}=
{\mu\over \left((R-1)\lambda_+\lambda_-\right)^{1\over R-2}}.}
Using the formula \ppbbmm\ again, we arrive at the following small
$\mu$ expansion:
\eqn\suscepty{\chi_0(\lambda,\mu)\equiv\partial_\mu^2
F_0(\lambda,\mu)= -{R\over 2-R}
\log\left((R-1)\lambda_+\lambda_-\right)+ {R\over
2-R}\sum_{n=1}^\infty{(-y)^n\over n!}  {\Gamma\left({n\over
2-R}\right)\over \ \Gamma\left({n\over 2-R}-n+1\right)}.}
The coefficient of $\mu^n/n!$ in \suscepty\ is the $n$ point
function of zero momentum tachyons in the Sine-Liouville background
\sineliouv. According to the conjecture of \FZZ,
for $R=3/2$ it should also be the $n$ point function
of these tachyons in the Euclidean black hole geometry.

To summarize, we find that for $1<R<2$, there is no obstruction to
turning on a large (real) Sine-Liouville coupling \conepert. There is a
critical point corresponding to $\mu=0$, $\lambda$ finite, which
appears to be stable and according to the conjecture described in
section 2 is equivalent (for $R=3/2$) to the black hole background
\cosetconf.

This raises an interesting question,\foot{D.K. thanks S. Shenker
and L. Susskind for a discussion of this issue.} what is the
interpretation of the Sine-Liouville background with $R\not=3/2$
in terms of the cigar geometry? Naively, changing the asymptotic
radius of the cigar creates a conical singularity at the tip, and
thus breaks conformal invariance. One natural proposal is that
changing the radius $R$ in the Sine-Liouville description corresponds
in the cigar theory to changing the asymptotic radius in the same
way (so the asymptotic geometries agree) and turning on a potential
like that in \conepert. This might lead to a change of the radius
of $x$ as a function of $\phi$, such that the conical singularity
at the tip is eliminated. It would be interesting to make this
proposal more precise, but we will not attempt that here.

In the next two sections we will construct a matrix model
describing two dimensional string theory with the worldsheet 
Lagrangian \conepert\ and use it to compute the
string partition function. We will in fact study
a more general problem corresponding to two dimensional
string theory in an arbitrary winding background,
\eqn\conepertt{
 L={1\over4\pi}[(\partial x)^2 +(\partial\phi)^2
 +2\hat R\phi+ \mu\phi e^{-2\phi}
  +
 \sum_{n\ne 0}  t_{n}  e^{(  |n|R-2)\phi}  
e^{i n R(x_L-x_R) }]. }
The couplings $t_n$ carry winding number $n$; the model
\conepert, \partm\ corresponds to the choice
\eqn\tlam{
t_{n }= \delta_{n,1}\lambda_{+} +\delta_{n, -1}\lambda_{-}.}
Of course, most of the couplings $t_n$ are in general irrelevant,
which is reflected in the fact that the corresponding
perturbations in \conepertt\ are singular at $\phi\to\infty$.
These couplings should be treated perturbatively, and the path
integral \conepertt\ should be thought of as the generating
functional of correlation functions of these operators, as a 
function of the relevant and marginal couplings.

\newsec{A matrix model for the  $c=1$  string
theory with vortices  }

In this section we will describe a matrix model for the $c=1$
string theory in the presence of vortices of arbitrary vorticity,
or in continuum language the theory described by the Lagrangian
\conepertt\ with $t_n\not=0$. We will 
show that the grand-canonical partition function of this matrix 
model is a $\tau$-function of the Toda chain hierarchy and, as 
such, it satisfies the Toda lattice equation. This equation can 
be used to compute the partition sum of the model to all orders 
in the string coupling (or equivalently to all orders in $1/M$ in 
the black hole background).

\subsec{\it The role of twisted boundary conditions in MQM}

The usual $c=1$ string \KAZMIG\ corresponds to the fixed point 
in which all $t_{n}$ in \conepertt\ vanish and the
scale is set by the cosmological constant $\mu$. The (perturbative)
dynamics is in this case well described by the singlet sector of the
$c=1$ matrix model and, as mentioned above, contains very few
degrees of freedom. The Euclidean problem with finite $R$
corresponds to finite temperature, $T=1/2\pi R$. When the temperature
is sufficiently low ($R>2$ in our normalizations \plr, \deltalr),
one can neglect the non-singlet states \GRKL. Above this critical
temperature ($R<2$) the vortices become important; studying the
Lagrangian \conepertt\ with $t_n\not=0$ is a useful way to incorporate
them.

To describe the c=1 string with vortex excitations \conepert\ we
will use the twisted MQM introduced in \BULKA.  Its canonical 
(fixed $N$) partition 
function is defined as a functional integral with respect to the 
one-dimensional $N\times N$ Hermitian matrix field $M_{jk}(x)$ %
\eqn\GENF{ \CZ_N(\Omega)
= \int_{M(2\pi R)=\Omega^{\dagger} M(0)\Omega} {\CD}M(x) e^{ - \tr
\int ^{2\pi R} _0 dx \left[ \hf (\p_x M)^2  +V(M)\right] }}
where $\Omega$ is an arbitrary unitary matrix. Note that the
path integral \GENF\ depends on $\Omega$ only through its
eigenvalues, $z_1=e^{i\theta_1}, \cdots, z_N=e^{i\theta_N}.$
The potential $V$ can be taken to be any function with the
appropriate critical behavior, \eg\
\eqn\vvmm{V(M) = \hf M^2 - {g\over 3\sqrt{N}}M^3.}
The projection onto the singlet sector is achieved by imposing 
the twisted boundary conditions \GENF\ and then integrating with 
respect to the twist matrix $\Omega$. As we review in Appendix A, 
the singlet sector can be reduced to a system of free fermions
corresponding to the eigenvalues $(y_1,\cdots,y_N)$ of the
matrix $M$, each compactified on the circle, $y_j(x+2\pi R)=y_j(x)$ 
and moving in the potential $V(y)$.  The introduction of vortex 
excitations is achieved by including higher  representations of 
$SU(N)$ \refs{\GRKL,\BULKA}. The eigenvalue dynamics in these
representations is more complicated.

To find the realization of the system \conepertt\ in MQM
we next take a closer look at the role of $\Omega$ in the
large $N$ limit. The $1/N$ expansion of the free energy
$$\CF_N(\Omega)= \log \CZ_N(\Omega)$$ 
of the $\Omega$-twisted 
matrix model \GENF\ can be expressed in terms of connected
planar graphs, embedded in the target-space circle.  Namely
\eqn\frenm{
\CF_N(\Omega)=
\sum_{h\ge 0} \CF_N^{(h)}\! (\Omega)\   N^{2-2h}, }
where $\CF_N^{(h)}\! (\Omega)$ is the contribution of planar
graphs with the topology of a sphere with $h$ handles.
We will use here the usual connection between planar graphs in MQM
and string worldsheets: each planar graph
$\Sigma$ is viewed as a discretized worldsheet whose vertices,
links and faces are labeled by $v$, $\ell$ and $f$,
correspondingly, and the functional integral over the worldsheet
fields $(x,\phi)$, where $\phi$ is the conformal (``Liouville'')
factor of the worldsheet metric, is discretized as $ \int \CD
x\CD\phi \ \to \ \sum_{\Sigma} \int \prod _v dx_v.$

\lref\bkt{V. L. Berezinski, JETP {\bf 34} (1972) 610;
J. M. Kosterlitz and D. J. Thouless, J. Phys.
{\bf C6} (1973) 1181; J. Villain, J. Phys. {\bf C36}
(1975) 581.}

The planar diagram expansion is done with respect to the trivial
(non-physical) vacuum, $M=0$.  The propagator of the matrix field
$M$ with twisted boundary conditions is
\eqn\PROP{\langle M_i^k(x) M_j^l(x')\rangle\equiv
G_{ij}^{kl}(x, x')= \sum_{m=
-\infty}^\infty e^{-|x -x'+2\pi R m|}(\Omega^{m} 
 )_{ij} (\Omega^{-m})^{kl}. }
To each link $\ell$ = $\langle vv'\rangle $ 
of the planar graph is associated a propagator
 $$x_v\
^{^i}_{_k}\!\!\!\!\!\!\!\ = \! = \! = \! =\!\!\!\! \!_{_{m}} \!
\!\!\!  = \! = \!  = \! = \!\!\!\! ^{^ j}_{_l}\ x_{v'}= 
G_{ij}^{ kl}(x_v, x_{v'}; m),$$
which is given by the $m$-th term in the expansion \PROP. 
This   gives rise to
  an integer-valued local field $ m= m_{ <\!vv'\!>}$ on the graph. 
The weight of each planar graph (see Fig. 2) is given by 
\eqn\weighT{
\CW(\Sigma) = N^{2-2h} g^{n_v}\ \prod_{f\in \Sigma}
{\tr \Omega  ^{w_{f}}\over N}
\exp\left(-\sum_{<vv'>} |x_v-x_{v'} +2\pi R m_{<\!vv'\!> } |\right),
}
where $n_v$ is the total number of vertices in the graph; the
integer field $w_f$  is related to the field $m_{ <\!vv'\!>}$ by
\eqn\mef{ w_f= \sum_{\ell\in \p f} m_{\ell}.} 
where the sum is taken along the links on the boundary of the
face $f$.

\bigskip
\vskip 5pt
    
    %
\hskip 40pt
\epsfbox{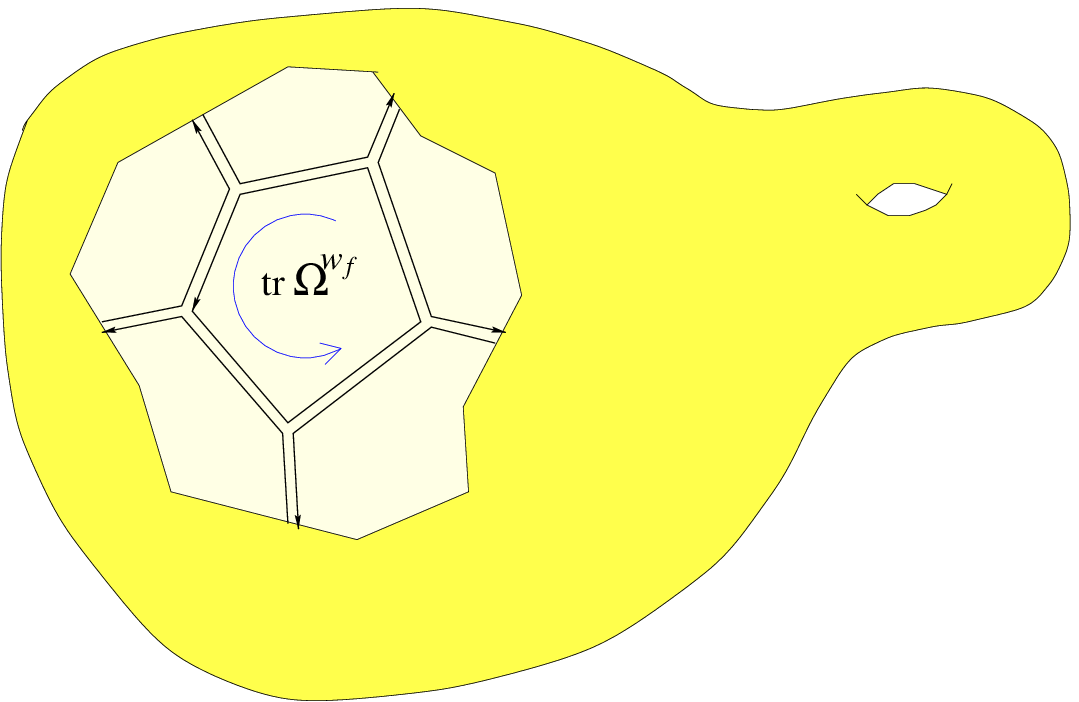}
\vskip 5pt
\centerline{Fig. 2:  A discretized worldsheet.}
\bigskip

Geometrically,  $w_f$ is the winding number
of the boundary $\p f$ of the face $f$ on the
discretized worldsheet around the
circle.  By definition this is the vorticity
associated with this face. Following standard
arguments \bkt,
the sum over the field $m_\ell$ can be split
into a sum over the vorticity field $w_f$ and
its gradient (vorticity-free) part.\foot{The latter
sum, corresponding to $m_{vv'}=m_v-m_{v'}$
for some $m_v,m_{v'}\in Z$, extends the
integration over $x_v$, which was originally
restricted to the interval $[0,2\pi R]$, to the
whole real axis.}
   
The dependence of a given planar graph on $\Omega$
is through the factor $\prod_{f\in \Sigma} \tr \Omega^{w_f}$;
hence the twist matrix can be used to introduce fugacities
of vortices with different vorticity.\foot{This generalization
of the arguments of \BULKA\ was suggested by P. Zinn-Justin.}
To study the effects of vortices on the worldsheet, one can
either analyze the dependence of the partition sum \GENF\ on
$\Omega$ (recall that in fact it depends only on the
eigenvalues of $\Omega$), or integrate with respect to the 
twist matrix with an appropriate measure. We will use the second 
alternative, which is technically preferable.

We start by showing 
that the singlet sector of the model is obtained by integrating 
over $\Omega$ with the standard left/right invariant (Haar) measure
$[d\Omega]$ normalized so that $\int_{_{U(N)}}[d\Omega]=1$.
An elegant way to evaluate the integral $\int_{_{U(N)}}[d\Omega]\
\tr\Omega^{n_1}\cdots\tr\Omega^{n_k}$  for $ n_1+\cdots+n_k<<N$ was
proposed in \DOUGLAS. It uses a Fock space representation of the
moments of the unitary matrix in terms of bosonic oscillator modes.
Replacing $\tr \Omega^n \to \alpha_{-n} + \bar \alpha_n$,
where $\{\alpha_n \}$ and $\{\bar \alpha_n \}$ satisfy the canonical
commutation relations $[\alpha_m, \alpha_n] =[\bar \alpha_m,\bar
\alpha_n]= m\delta_{m+n, 0}$, $[\alpha_m, \bar\alpha_n]=0$,
the above integral can be thought of
as the expectation value with respect to the Fock vacuum $\alpha_n |
0\rangle = \bar \alpha_n | 0\rangle =0$. This leads to:
\eqn\intUN{\eqalign{
 \int_{_{U(N)}}[d\Omega]\ \ 
\tr\Omega^{n}\ & = N\delta_{n,0}\ ,\cr
\int_{_{U(N)}}[d\Omega]\ \
\tr\Omega^{m}\ \tr\Omega^{n}&=
|m|\delta_{m+n, 0}\ ,  \ {\rm etc}.}}
Using these equations it becomes clear that the sum over
all Feynman graphs \weighT\ with a given set of vortices
$w_f$ corresponds in the continuum language to the string
worldsheet path integral with an insertion of the
corresponding winding modes, \tachp:
\eqn\coromt{{\rm tr}\Omega^n\leftrightarrow
\CT_{nR}.}
The $\delta$-functions in eq. \intUN\ impose winding
number conservation.

We would like to analyze the behavior of the partition sum
\eqn\ppss{\int [d\Omega]\ \CZ_N(\Omega)}
in the double scaling limit. By using \weighT\ and the first line
of \intUN\ we see that each Feynman diagram with $w_f=0$ for all 
faces contributes a term that goes like $N^{2-2h}$ (where $h$ is 
the genus of the discretized Riemann surface). The continuum limit 
is achieved in the matrix model by tuning the coupling $g$ in \vvmm\ 
to a critical point $g\to g_c$ where large Feynman diagrams dominate 
the path integral, and considering the double scaling limit $N\to\infty$,
$g\to g_c$ with $N(g_c-g)$ held fixed.\foot{up to certain logarithmic
corrections, which are not essential here.}
The sum over diagrams corresponding to large worldsheets 
with $w_f=0$ scales like $\left[(g_c-g)N\right]^{2-2h}\sim \mu^{2-2h}$
in this limit.  

Feynman diagrams \weighT\ with $w_f\not=0$ do not
contribute to the partition
function in the double scaling (continuum) limit. In addition to the
factor $\mu^{2-2h}$ discussed above, each insertion of a vortex is
suppressed by $1/N$, and
the sum over graphs gives rise to a factor $(g-g_c)^{-\alpha}$ where
$1-\alpha=|w_f|R/2$ is the KPZ scaling dimension of the winding mode
$\CT_{w_fR}$ \tachp . Since $\alpha<1$, the contributions of surfaces
with vortices go to zero in the double scaling limit.

Therefore, we conclude that the partition sum \ppss\
corresponds in the continuum limit to a sum over discretized
worldsheets with no vortices, as claimed above.

\subsec{Introducing vortex couplings in the  matrix model}

To enhance the contribution of vortices in the double scaling
limit we replace the left-right invariant Haar measure
$[d\Omega]$ on the $U(N)$ group manifold by
\eqn\measom{
 [D\Omega]_{\{\lambda \}}= [D\Omega]\  \exp\left( \sum_{n\in\Z} \lambda
_n \tr \Omega^n\right) }
which is invariant only with respect to internal automorphisms
$\Omega\to U^{-1} \Omega U$. The corresponding partition
function is
\eqn\AVTW{
\CZ_N[ \lambda]= \int [D\Omega]_{\{\lambda \}}\  \CZ_N(\Omega
).}
Applying the integration rules \intUN\ we see that in addition
to the contractions discussed above, which vanish in the double
scaling limit, we now have new terms where we use the second line
of \intUN\ to contract a ${\rm tr} \Omega^n$ from the
exponent in \measom\ with one of the factors ${\rm tr} \Omega^{w_f}$
in a Feynman graph \weighT. This leads to a factor of $\lambda _{w_f}$
for each face with vorticity $w_f$. Therefore, the coupling constants
$\lambda_n$ can be thought of as fugacities of vortices with
vorticity $n$ ($n=\pm 1, \pm 2, ...)$. 

One can also use the rules \intUN\ to contract the traces
in the exponent in \measom\ among themselves. This gives rise to a
multiplicative factor $\exp\sum_{n\ge 1}n \lambda_n \lambda_{-n
}$ in the partition function, which is large but ``non-universal''
in the language of non-critical string theory.

To summarize, the free energy 
\eqn\ccffnn{\CF_N[ \lambda]= \log \CZ_N[\lambda]}
is given in the double scaling limit by a sum over all discretized
worldsheets containing vortices, where the chemical potential of
vortices with vorticity $n$ is $\lambda_n$ (up to a multiplicative
factor which we will implicitly determine below).  We conclude that 
the matrix model \AVTW\ gives a discretization of the string theory 
\conepertt.

\subsec{\it Integrating  over the twist
angles in the double scaling limit}

The twisted partition function \AVTW\ can be calculated
in the double scaling limit, where the matrix model
simplifies drastically.  As is familiar from the usual
treatment of two dimensional string theory, in this limit
the dynamics takes place near the top of the matrix potential
$V(M)$ \vvmm, and is dominated by large planar diagrams. Thus,
to study the theory in the double scaling limit we can focus on
the vicinity of the quadratic maximum of \vvmm\ and replace the
potential by $V(M)=-M^2/2$
(after shifting $M\to M-\sqrt{N}/g$). The behavior of the potential far
from the top enters as a UV cutoff, which is usually taken to be
a wall placed at a distance $\Lambda\sim \sqrt{N}/g$ from the top
\refs{\BRKA ,\PARISI ,\GRMI ,\GIZI,\KAZREV,\KLEBANOV,\ginsmoore}. The
twisted  partition function \GENF\ now reads
\eqn\INVOS{\CZ_N(\Omega)
= \int_{M(2\pi R )=\Omega^{\dagger} M(0)\Omega}
{\CD}M  e^{  \tr \int ^{2\pi R} _0 dx \left(- \hf (\p_x
 M)^2 +\hf M^2  \right) },}
where the functional  measure $ {\CD}M$ is regularized
as discussed above.

For generic  twist matrix $\Omega$  we actually do not need to
regularize the measure because the matrix integral is convergent.
In the compactified theory, the cutoff has no effect when the
compactification radius is smaller than half the self-dual radius
($ R<1/2$ in our units); we review  this argument of \BULKA\
in Appendix B. The Gaussian integration gives then a simple
determinant depending only on the eigenvalues $z_1, ..., z_N$
of the twist matrix $\Omega$, which are phases,
$z_j=\exp(i\theta_j)$:
\eqn\twpa{ \CZ_{N} ( \Omega ) = \prod_{j, k=1}^N {1\over |z_{j
  } q^{1/2}- z_{k} q^{-1/2}| } , }
where
\eqn\QQQQ{
q = e^{2\pi i R}.  }
 Note that the partition sum \twpa\ depends on $R$ only through $q$
\QQQQ, and is also symmetric under $q\to q^{-1}$; the integral over
the twist matrix \AVTW\ preserves this property. Naively, this means
that $R$ can be restricted to the interval $0<R<\hf$.  Later we will
transform to the grand canonical ensemble, in which we will see that
the symmetry is broken due to the appearance of certain divergences.

The next step in evaluating the partition sum is the integration over
$\Omega$ in \AVTW. Due to the symmetry of the measure with respect to
internal automorphisms, the volume of the diagonal $U(N)$ group
factors out and the $\Omega$ integral reduces to an integral over its
eigenvalues $z_1, ...  z_N$:
\eqn\ommeasure{
\int [D\Omega]_{\{\lambda\}}\ \CZ_{N} ( \Omega )= {1\over N!}
\prod_{k=1}^N \oint {dz_k\over 2\pi i}\ e^{2u(z_k)}\
|\Delta(z)|^2\ \CZ_{N} ( \Omega ) }
where $\Delta(z)$ is the Vandermonde determinant (see
\eg\ Appendix A) and we introduced the potential
\eqn\potu{u(z) = \hf \sum_{ n}\lambda_n z^n.}
Plugging \twpa\ in \ommeasure, we find that 
\eqn\canpart{\CZ_N[\lambda]={1 \over N!}\oint \prod_{k=1}^{N} 
\ {dz_k\over 2\pi i z_k} \ {e^{ 2 u(z_k)}
\over ( q^{1/2}-
q^{-1/2} ) } \ \prod_{ j \ne j' } ^{N} {z_j \ -\ z_{j'} \over q^{1/2}
z_j - q^{-1/2} z_{j'}}.}
Another useful representation of $\CZ_N$, obtained by using
the Cauchy identity 
\eqn\cauid{{\Delta(a)\Delta(b)\over\prod_{i,j}(a_i-b_j)}=
\det_{(i,j)} \left({1\over a_i-b_j}\right) }
is
\eqn\cccddd{\CZ_N[\lambda]= 
\prod_{k=1}^N \oint {dz_k\over 
2\pi i }
\det_{(i,j)}\left(  {\exp\left[u( z_{i}) +u( z_{j }) \right]
\over z_{i } q^{1/2}- z_{j} q^{-1/2}}\right).}
A natural way to avoid the ambiguity in the contour integration in
\cccddd\ is to add a small imaginary part to the radius $R$ and
integrate over the $\{z_k\}$ along the unit circle.\foot{Note that
with this prescription we do not obtain a doubling of the degrees of
freedom, typical for matrix models with  symmetric potentials.}

 The partition
function can be calculated by residues for any $N$ and $t_k$, but the
formulae get more and more complicated when $N$ grows. Furthermore,
the analytical structure of $\CZ_N$ makes it difficult to formulate a
well defined $1/N$ expansion.

\subsec{The grand canonical partition function as a 
$\tau$-function of the Toda chain hierarchy}
 
The abovementioned problems can be overcome by considering
instead of \canpart\ the grand canonical partition function,
where the cosmological constant $\mu$ becomes the
chemical potential. Below we will show that  the latter
 is a $\tau$-function of the infinite Toda
chain hierarchy \UT\ and as such it satisfies the Toda equation (with
appropriate boundary conditions), which plays the same role as the
Painlev\'e equation of the KdV hierarchy in the $c=0$ string theory.
It will be convenient to rescale the vortex couplings and define new
couplings
\eqn\teens{ t_n = { \lambda_{n} \over q^{n/2} -q^{-n/2} },}
which are commonly used in the literature
on  $\tau$-functions.

The grand canonical partition function is defined by
\eqn\gkpf{ \CZ_\mu[\lambda] =\sum _{N=0}^{\infty} e^{2\pi
R \mu N}\  \CZ_N[\lambda].}
Using \cccddd, one can show that the grand canonical partition 
function \gkpf\ can be represented as a Fredholm determinant
\eqn\Zdet{ \CZ_{\mu}= {\rm Det} (1+e^{ \mu 2\pi R} \hat K),
}
where the integral operator $\hat K$
 is defined as
$$(\hat Kf)(z)=-\oint {dz'\over 2\pi i}\ {e^{u(z) +u(z')}
\over q^{1/2} z- q^{-1/2} z'}\ f(z'). $$
In Appendix $C$, the partition function \Zdet\ is identified
as a solitonic $\tau$-function of the Toda chain hierarchy.\foot{
Usually a solitonic $\tau$-function is defined by a finite sum instead
of an integral. Our partition function is a particular case of the
continuum limit (infinitely many solitons) of the soliton solution
referred to as ``general solution'' in ref. \JM .} Explicitly the
$\tau$-function with charge $l$ is defined as
\eqn\tauf{\tau_l[t] = e^{-\sum _n nt_{n}t_{-n} } 
\sum_{N=0}^{\infty} (q^{ l}
\ e^{2\pi R \mu } )^{N}  \CZ_N[t]=e^{-\sum _n nt_{n}t_{-n}}
\CZ_{\mu+il}[t]}
The grand canonical partition function \gkpf\ is obtained by
taking $l=0$. The $\tau$-function with charge $l$ can be obtained 
from \gkpf\ by replacing $\mu\to \mu + i l$.

\subsec{Toda equation for the free energy}

The $\tau$ functions $\tau_l$ satisfy a hierarchy of Toda lattice
equations which are differential equations in $\{t_n\}$ and finite 
difference equations in $l$. This infinite set of equations is generated
by the corresponding Hirota identity, and reflects the $Gl(\infty)$ 
symmetry of the system. This symmetry can be exhibited by 
representing the system  in terms of chiral free fermions. 
The lowest equation from the Toda chain  hierarchy is 
\eqn\TodaDD{\tau_{l} (\p_{+}\p_{- }  \tau_{l} )- (\p_{+} \tau_{l})
(\p_{-} \tau_{l} ) +\tau_{l+1}\tau_{l-1} =0, } 
where we denoted
$\p_{\pm} = \p/\p\l_{\pm}$ and $\l_{\pm } = t_{\pm 1}=\pm{\l_{\pm 1
}\over 2i\sin(\pi R)}$.

As mentioned above, the $\tau$-function
depends on the cosmological constant $\mu$ and the charge $l$ via
the combination $\mu +il$
\eqn\tred{  \tau_l( \mu )= \tau_{0} ( \mu +i  l).}    
We will restrict ourselves to the case \tlam\
(note that the constants $\lambda_\pm$ are related to
$\l_{\pm1}$ by \teens), for which
\eqn\relfff{\tau_{0}(\mu) =e^{-\lambda_+\lambda_-}\CZ_{\mu}(\lambda).} 
Winding number conservation implies that the $\tau$-function depends only 
on the product $\l^{2}= \l_{+}\l_{-} $, so that
$$
  \p_+   \p_-
  = {1\over 4  }  \l^{-1}  \p_\l \l\p_\l , \qquad
  \l=\sqrt{\l_{+}\l_{- }}  .
  $$

The free energy $\CF(\l,\mu) =  \log \CZ_{\mu}(\l)$
of the matrix model satisfies
\eqn\todisc{
\p_{+}\p_{-}
 \CF(\l,\mu) +\exp \left[ \CF(\l,\mu+i) + \CF(\l,\mu-i)-2 \CF(\l,\mu) \right]
=1.  }
In the double scaling limit $\CF(\l, \mu)$ is assumed to be a smooth function 
of $\mu$. Thus we can rewrite \todisc\ as
\eqn\TodaD{
\p_{+}\p_{-} \CF(\l,\mu)
+\exp\left(- 4\sin^2\left({1\over 2} {\p \over \p\mu}\right)
\CF(\l,\mu)\right)=1.}
Differentiating twice w.r.t. $\mu$, we get an equation for the
susceptibility  $\chi(\l, \mu) = \p_\mu^2 F(\l,\mu)$:
\eqn\TodaS{\p_{+}\p_{-}
    \chi(\l, \mu) + \p_\mu^2 \ 
\exp\left[- \left({\sin (\hf \p_{\mu})\over \hf\p_{\mu} }\right)^{2} 
\chi(\l, \mu)\right] =0. }
The operator in the exponent is to be understood as a power series in
$\p_{\mu}$:
\eqn\DOP{ {\sin (\hf \p_{\mu} )\over \hf\p_{\mu} } =1+\sum_{n=1}
^{\infty} (-)^{n} {2^{-2n} \over (2n+1)!}\p_{\mu} ^{2n} .}  
Note that the differential equations \TodaD, \TodaS\ do not depend on
$R$ explicitly. However, to solve them we need to supply boundary
conditions. A convenient choice is to specify the free energy $\CF$
at $\lambda=0$, and use the differential equations to solve 
for the $\lambda$ dependence. Since $\lambda=0$ corresponds to the 
standard $c=1$ matrix model, $\CF(\mu)\equiv \CF(\lambda,\mu)|_{\lambda=0}$ 
is known. We next 
review its form.

\subsec{Boundary conditions}  

The $c=1$ string theory without vortices corresponds to $u(\Omega)=0$ (see
\potu). The integral in \cccddd\ can be evaluated by
residues and gives %
\eqn\PFO{ \CZ_{N} = q^{N^2/2}\prod_{n=1}^N {1\over 1-q^n} .}
This formula makes sense for the usual matrix harmonic oscillator, but
for the inverted oscillator it requires an analytic continuation.
Fortunately, the corresponding 
grand canonical partition function \gkpf\ in the absence of
vortices (all $t_m=0$) can be expressed as a partition function of
fermions in terms of the density $\rho(E)$ of the energy levels of 
the inverted harmonic oscillator (see \GRKL).  The (universal 
part of the) grand canonical free energy of the $c=1$ string theory
compactified on a circle of radius $R$ has a well defined (asymptotic)
$1/\mu$ expansion:
\eqn\FrenO{\eqalign{ \CF(\mu)& = \log \CZ_{ \mu}[0] \cr 
&=\hf \int _{-\infty}^{\infty} dE\rho(E) 
\log\left(1+ e^{-2\pi R(\mu-E)}\right)\cr
&= {1\over 2\pi} \int _{-\infty}^{\infty} dE\sum_{k=0}^{\infty} {k+\hf
\over E^{2}+(k+\hf)^{2}} \log(1+e^{-2\pi R(\mu-E)})\cr
&  = -  {R\over2}\mu^2 \log \mu -
{1\over24}\big(R + {1\over R} \big)\log \mu +
R\sum_{k=2}^\infty \mu^{-2(k-1)} f_k(R)+O(e^{-2\pi\mu}),
}}
where we introduced the polynomials in ${1\over R}$
\eqn\POLYNR{ f_k(R)=
(2k-3)! \  2^{-2k} \sum_{n=0}^{k} 
\left({1\over R }\right)^{2n}
{(2^{2(k-n)}-2)(2^{2n}-2)\vert B_{2(k-n)} B_{2n}\vert \over
[2(k-n)]![2n]!}}
and $B_m$ are Bernoulli numbers.
The partition function \FrenO\ has a T-duality symmetry, 
$R \to {1\over R }$, $\mu \to R \mu$.  This symmetry
is broken in the presence of vortices (\ie\ for $\l\ne 0$).

\subsec{Canonical versus Grand Canonical partition sum}

The sum over worldsheets in the continuum string theory is
described in the matrix model by the $1/N$ expansion of
the canonical free energy $\CF_N=\log\CZ_N$. The formalism
described in the previous subsections allows one to compute 
the grand canonical free energy $\CF(\mu)=\log\CZ_\mu$, which
is related to $\CF_N$ via 
\eqn\CANGC{  \exp\CF_N=\oint {d\mu\over 2\pi i} \  \exp[2\pi R \mu N+ 
\CF(\mu)]
}
where the integration contour encircles the point $e^{-2\pi R\mu}=0$. 

As is well known \refs{\KAZREV,\KLEBANOV,\ginsmoore}, the $1/N$
expansion of $\CF_N(\lambda)$ can be rearranged as a $1/\mu_0$ 
expansion, with $\mu_0$ defined by $\mu_0\log(\mu_0/\Lambda)=N(g_c-g)$. 
The grand canonical free energy $\CF(\lambda=0,\mu)$ has a $1/\mu$ 
expansion given by \FrenO. Similarly, in section 5 we will show
that $\CF(\lambda,\mu)$ has a $1/\mu$ expansion which is obtained
from \TodaD. One can show that in the double scaling limit, 
the $1/\mu_0$ expansion of $\CF_N$ coincides with the $1/\mu$
expansion of $\CF(\mu)$, up to a flip of the sign\foot{This sign 
flip is a standard feature of the Legendre transform.} of the leading
(spherical) contribution to $\CF$ (see \BULKA, revised version, section 6). 

In the next section we will construct the $1/\mu$ expansion of
the grand canonical free energy, and use the above discussion
to identify it with the genus expansion of the partition sum 
of the continuum theory.

\newsec{Some results on the genus expansion of the free energy}

The purpose of this section is to solve the differential equations
\TodaD, \TodaS\ with the boundary conditions \FrenO. We have not
found an exact solution of these equations, but as we will see below,
one can solve them iteratively, in a genus expansion,
\eqn\freeenh{\CF(\l,\mu)=\sum_{h=0}^{\infty}\CF_h(\l, \mu).}
The genus $h$ partition sum $\CF_h$ must, by KPZ-DDK scaling, have
the form\foot{The fact that the r.h.s. of                
\TodaD\  is one, leads to an additive contribution $\lambda_+\lambda_-$ to
$\CF$. This contribution is non-universal and will be dropped below.}  
\eqn\ccffhh{\CF_h(\l,\mu)=-\delta_{h,0}{R\over 2}\mu^2\log\mu 
-{\delta_{h,1}\over24}(R+{1\over R})\log\mu+\mu^{2-2h}A_h(z),}  
where $z$ is defined in \defzz. The
boundary conditions \FrenO\ determine $A_h(z)$ at $z=0$. 
Plugging the  ansatz \ccffhh\ into \TodaD, \freeenh\ one can solve for
$A_h(z)$.  It is remarkable that the Toda equations allow such an ansatz at
all;  this is due to their conformal symmetry. 

At tree level one finds an ODE for $A_0(z)$ which together with the 
boundary conditions fixes it uniquely. At genus $h$ one finds a
differential equation that mixes $A_h(z)$ with $A_{h'}(z)$ with
$h'<h$. These equations can be used to determine all $A_h$ iteratively
given $A_h(0)$. 

We will actually perform the calculations by transforming from the
variables $(\mu, z)$ to $(\lambda, y)$, where $y$ is defined in 
\defyy. In these variables the genus expansion is an expansion
in inverse powers of $\lambda^{4\over 2-R}$, and the boundary
conditions are specified at $y=\infty$. The genus $h$ free energy
in \freeenh\ takes the form 
\eqn\PARTFU{\eqalign{
\CF_0(\l,\mu)& = (\eta \l^{2\over 2-R})^2 \left(- {R\over 2-R}
y^2\log(\l\sqrt{R-1}) + f_0(y)\right)\cr \CF_1(\l,\mu)& = C\log\l
+f_1(y)\cr \CF_h(\l, \mu)& = (\eta \l ^{2\over 2-R}) ^{2-2h}
f_h(y)\qquad (h\ge 2), }}
where, as in \defyy,  $$y =z^{-{1\over 2-R}}
={\mu\over \l^{2/(2-R)}\k}, \qquad \k=(R-1)^{1/(2-R)}.$$
The genus expansion corresponds to fixed $y$ with $\mu,\l\to\infty$. 
The logarithmic term in $F_1$ is a zero mode of the Toda
equation \todisc. The constant $C$ will be determined later.

We note in passing that $C$ is a special case of a large class of
zero modes of the Toda equation \TodaD, whose general form is
$$
\Delta \CF=   \sum_{n=0}^\infty
(C_n+D_n  \log \lambda  )\  e^{ -2n\pi\mu}.
$$
For $n\not=0$ these terms are non-perturbative in the genus expansion.
The $\{C_n\}$ are nothing but the non-perturbative terms in the
standard $c=1$ string theory \FrenO. Their presence is related \SHENKER\ 
to the $\sim( 2h)!$ divergence of the coefficients $f_h$ in the perturbative
expansion \PARTFU. The status of non-perturbative terms in two dimensional
string theory is generally unclear. Here we are interested in the perturbative
expansion  of the free energy and will therefore drop all exponential
corrections.

Differentiating \PARTFU\ twice w.r.t $\mu$ we find the susceptibility,
\eqn\SEX{\eqalign{ \chi(\l,\mu)&=\sum_h 
\chi_h(\l,\mu)\cr
\chi_0(\l,\mu)&=-
{2R\over 2-R} \log \left(\l\sqrt{R-1}\right) + X_0(y),\cr
\chi_h(\l,\mu)&= \left(\eta\l^{2\over 2-R}\right)^{-2h}  X_h(y)
\qquad(h\ge 1), }}
with $f_h$ and $X_h$ related as
\eqn\CHIFS{ \p_y^2 f_h(y)= X_h(y) .  }

\subsec{  Partition function on the sphere  }

Plugging \SEX\ into \TodaS\ and retaining in the spherical approximation
only the $h=0$ term (and only the leading term in \DOP ) we obtain
the following ODE defining the string susceptibility $X_0(y)$
\eqn\SEQ{ \a\left(y\p_y\right)^2 X_0+\p_y^2 e^{- X_0}=0  }
where $\a=(R-1)/(R-2)^2$. The solution is
\eqn\SFSOL{y= e^{-{1\over R} X_0} - e^{ {1-R\over R} X_0},}
reproducing Moore's result \finsuscep -- \xxyy .  This expression
can be checked by substitution or found by a direct solution of 
\SEQ\ (see Appendix D). The integration constant is fixed by
comparing to the leading term in \FrenO.

Using \CHIFS\ and integrating $X_0$ twice we obtain the following
parametrization of the partition function on the sphere:
\eqn\PARTO{\eqalign{  &\CF_0(\l,\mu)=
 -\hf \mu^2\left( {2R\over 2-R}\log (\l\sqrt{R-1})-
 X_0 \right) \cr   
& +\eta^2\l^{4/(2-R)}\left({3\over 4} {R\over R-1} 
e^{-2{R-1\over R} X_0}+{3\over 4}R e^{- {2\over R}X_0}-
{R^2-R+1\over R-1} e^{- X_0} \right)     } }
with $X_0(y)$ defined from \SFSOL. 

Consider the limits $y\to 0,\infty$ in \PARTO.  In the
limit $y\to\infty$ ($\mu\to\infty$ or $\l\to0$) \PARTO\ 
reproduces the known asymptotics of the $c=1$ string theory 
perturbed by
vortices:
\eqn\wwpp{\CF_0\simeq - \hf R\mu^2\log \mu - \lambda^2\mu^R+
O(\lambda^4\mu^{2R-2}), \ \ \ \mu\to\infty}
whereas in the Sine Liouville limit $\l\to \infty$ with $\mu$ fixed we
obtain the asymptotics
\eqn\BHLIM{ \CF_0\simeq -A \l^{4/(2-R)} -
B\mu\l^{2/(2-R)} +O(\mu^2),
\quad \l\to\infty. }
where $A= {1\over 4}(2-R)^2 (R-1)^{R/(2-R)}$ and $B=
(R-1)^{1/(2-R)}{(2-R)^3\over R(R-1)}$.
\foot{ Note that a more natural parameter would be 
$\tilde\l=\sqrt{R-1}\l$; indeed, loking back at \suscept,
\defzz\ we see that the dependence of the partition sum
on $\lambda$ is through $\tilde\lambda$. In terms of $\tilde
\lambda$ one finds that  
$\CF_0\simeq -{1\over 4\alpha }\tilde\lambda^{4/(2-R)}$.}  Note that
the dependence of the partition sum on $\lambda$ is in agreement with
the KPZ-DDK analysis in Sine-Liouville theory.  In particular, for the
black hole radius $R=3/2$, we find $\CF_0\sim \l^8$, in agreement with
\pertz.  It is interesting that the partition sum on the
sphere goes like an even integer power of $\lambda$. This
is a non-universal contribution to the partition sum of Sine Liouville
theory, which corresponds to condensation of vortices.
Recall that $\lambda$ is the fugacity of
vortices, and a finite integer power corresponds to a contribution of
a finite number of vortices and antivortices. Hence it should be
dropped from the physical answer.  As we shall see in section
6, this has an important physical consequence: the free energy
of the two dimensional black hole vanishes in the tree
approximation, in agreement with the results of \GP.

Not also that the coefficient $A$ is complex for $R<1$, in agreement with
our earlier analysis -- the imaginary part for $R<1$ signals the
instability of Sine-Liouville (with real $\lambda$) in this range.

\subsec{ Linear equations for any genus}

Substituting \PARTFU\ into \TodaS\ 
we obtain the following ODE for $f_h$
\eqn\ANYH{ \a (y\p_y+2h-2)^2 f_h-e^{-X_0(y)} \p_y^2f_h=
\CH( f_0,\cdots, f_{h-1}).   }
Here $X_0(y)$ is defined by \SFSOL\  and the function
$\CH( f_0,\cdots, f_{h-1})$  is defined as
\eqn\FORCE{ \eqalign{ &\CH(f_0,\cdots, f_{h-1})=  \cr
& -\left[(\sqrt{R-1}\l)^{{-{4-4h\over 2-R}}}
\exp\left(- 4\sin^2 (\hf \p_{\mu}) \
\sum_{k=0}^{h-1} (\sqrt{R-1}\l)^{-4k/(2-R)} f_k\right)\right]_0,  }}
where by $[\cdots]_0$ we denoted the constant term in
the limit $\l,\mu\to\infty$ with $y$ fixed.

We see that the equations for $\{f_h\}$ form a triangular
system of linear second order ODE's with $y$ dependent coefficients.
We do not know how to solve them in general; in the next subsection
we will describe the solution for the case of genus one (the torus),
where the equation simplifies.

\subsec{Partition function on the torus}


Eqs. \ANYH\ and \FORCE\ lead to the following first
order linear inhomogeneous ODE for the torus partition function $f_1(y)$
\eqn\FIRST{ (\a y^2-e^{-X_0})f_1''+\a y f_1'=
-{1\over 12} X_0''e^{-X_0}.}
Note that \FIRST\ is a first order differential equation for $f_1'$.
Integrating this equation and using the boundary condition at large 
$\mu$ \FrenO\ to fix the integration constant (see Appendix D) we 
obtain (up to a non-universal constant) 
\eqn\GENZ{ \CF_1(\l,\mu)={R+R^{-1}\over 24}
\left(R^{-1}X_0-{2\over 2-R}\log (\l\sqrt{R-1})\right)-
{1\over 24}\log \left(1-(R-1)e^{{2-R\over R} X_0}\right) }
with $X_0(y)$ defined by \SFSOL .
 In the limit $\mu\to\infty$ with $\l$ fixed we reproduce from \GENZ\
the known asymptotics of the $c=1$ string theory perturbed by
vortices: 
\eqn\TOREX{ \CF_1\simeq -{R+R^{-1}\over 24}\log \mu + {R^2-1\over 24
R}\l^2\mu^{R-2} + \cdots, \ \ \quad (\mu\to\infty) }
whereas in the Sine-Liouville limit $\l\to \infty$ and $\mu$ fixed we
obtain the asymptotics
\eqn\TORLIM{\CF_1\simeq -{R+R^{-1}\over 12(2-R)}\log (\l\sqrt{R-1})  
-{R-1\over 24 (2-R)}{\mu\over \left((R-1)\l^2\right)^{1/(2-R)}}+
\cdots, \quad  
{(\l\to\infty). }}
%

The expansion \TOREX\ of the solution \GENZ\ can be compared with the
explicit expressions for genus one tachyon correlation functions in
$c=1$ string theory obtained in \HK. Since \HK\ studied momentum
modes, to compare one has to perform a T-duality transformation
$\tilde R = 1/R$, $\tilde \mu = R\mu$. After this redefinition, the
results of \HK\ read:
\eqn\HSKU{ \langle \CT_{R}^{n} \CT_{-R}^{n} \rangle_{{h=1,R}} =
-{1\over R}\ {n!\over 24}\ [(1-R)\mu^{R-2} ]^{n}  
\left( (-)^{n} f_{n}(R) +g_{n}R^2\right)} 
where
 $$
 g_{n}(R)= {\Gamma(n(2-R) )\over \Gamma(n(1-R) +1)}$$
 and
 $$\eqalign{&f_{1} (R) = R^{2}-R -1,\cr
  & f_{2} (R) = 3R^{3}-8R^{2} +3R+3,\cr
  & f_{3} (R) = 17R^{4}- 72R ^{3}+ 90 R^{2}-17R -20  ,\cr
  & {\rm etc.}\cr}
 $$
For low $n$ one finds 
\eqn\coefst{
\langle\CT_{+}  \CT_{-}  \rangle_{h=1} = {R^{2}-1\over 24 
R}\mu^{R-2}   ,\ \ \ 
\langle\CT_{+}^{2} \CT_{-} ^{2} \rangle_{h=1} =
-{1\over 12 R} (1-R)^{2}(R^{3}-5R^2 +3R +3) \mu^{ 2R-4}.   }
This is exactly what one obtains by expanding the solution \GENZ.

\subsec{Comparison to continuum calculations}

Most of the matrix model results described above are difficult
to derive using continuum methods. In this subsection we will
discuss two cases in which the matrix model predictions can actually
be obtained directly in the continuum formulation; the non-vanishing
of the partition sum on the sphere and the precise value of the
partition sum on the torus.

Consider first the genus zero case. In the matrix model we found
a non-zero partition sum given by \BHLIM\ (with $R=3/2$ for the
Euclidean black hole case). In the continuum analysis on the 
cigar\foot{A similar analysis can be performed directly in Sine-Liouville
theory.} this can be reproduced as follows.

In string theory the partition sum on the sphere is usually said 
to vanish, due to the volume of the Conformal Killing Group (CKG)
of the sphere, $SL(2,C)$. If the target space is non-compact, the 
partition sum is actually proportional to $V/{\rm vol}(SL(2,C))$  
where $V$ is the divergent volume of spacetime. Thus, at first sight  
the partition sum is $\infty/\infty$, \ie\ it seems to be
ill defined. However,
in   most situations one is actually interested in the partition sum per  
unit volume. E.g. if the vacuum is translationally invariant in 
the non-compact directions, the partition sum per unit volume is the  
Lagrangian density in this vacuum (the classical cosmological constant),  
and it vanishes due to the volume of the CKG. 

In the Euclidean cigar background $H_3^+/U(1)$, the above discussion has  
to be  reexamined. There is no reason to divide by the
volume of the cigar since the background is not translationally invariant
in $\phi$, and in any case for comparison with the matrix model we are
interested in the total partition sum and not the partition sum per unit
volume. The ratio of the volume of the cigar to the volume of the CKG is
finite in this case. The volume of $H_3^+$ contains precisely the same
kind of divergence as that of the CKG. Since the volume of the 
$U(1)$ in \cosetconf\ is finite, we conclude that the partition 
sum of string theory in the cigar background is non-zero (and proportional
to the mass of the black hole
$M$). We will  not attempt to compute its precise value.

Moving on to the torus, we would like to derive the result \TORLIM\
directly in the continuum formalism. We will describe the calculation
in the Sine-Liouville language, to study the $R$ dependence. Similar
considerations can be used to perform the calculation on the cigar.

The torus partition sum in the background \sineliouv\ is an example
of a bulk amplitude in the language of section 2. By performing the 
integral over the zero mode of $\phi$ as in eq. \bulkam\ and using
\regbulk\ one finds 
\eqn\zttt{
Z_{\rm torus}=-{1\over 2-R}\log{\lambda\over\Lambda}
\int {d^2\tau\over\tau_2^2}Z_1(\tau,\bar\tau),}
where $Z_1(\tau,\bar\tau)$ is the torus partition sum of the theory with
$\lambda=0$, \ie\ string theory on an infinite cylinder of radius $R$.
$Z_1$ can be computed by using free field techniques; the linear
dilaton term in \sineliouv\ can be ignored in this case since the 
curvature of the torus vanishes. Equation \zttt\ has a simple
interpretation: the bulk of the contribution to the torus partition
sum comes from the region far from the Sine-Liouville wall. It can
be computed by cutting off the infinite region very far from the wall,
by restricting to $\phi\le\phi_{UV}$. As indicated by the notation, this
can be thought of as adding a UV cutoff to the theory. 

The Sine-Liouville wall acts as an effective IR cutoff, restricting
\eqn\rreett{\lambda e^{-(2-R)\phi}\le 1,\;\;{\rm or}\;\;
\phi\ge{1\over 2-R}\log\lambda.}
Thus, the length of the cylinder is
\eqn\vvff{L_\phi\simeq\phi_{UV}-{1\over 2-R}\log\lambda.}

The prefactor of the modular intgeral in \zttt\ is hence nothing but
the (universal part of the) length of the cylinder, and the integral
computes the free energy of perturbative string modes (tachyons)
living on the cylinder.

This makes it clear that the coefficient of $-{1\over 2-R}\log\lambda$
in \zttt\ must be the same as the coefficient of $-{1\over2}\log\mu$
in the torus partition sum in standard $c=1$ string theory, which has been
computed by performing the integral \zttt\ (see \eg\ \KLEBANOV). As we see 
in \FrenO, this coefficient is $(R+{1\over R})/12$, in agreement with \TORLIM.

\newsec{Thermodynamics of two dimensional string theory}

Our results from the previous sections can be used to study
the thermodynamics of two dimensional string theory.
We start by reviewing the low temperature thermodynamics
(corresponding to $R>2$ in the notations of the previous
sections), and then describe the physics for temperatures slightly
below and slightly above the Hagedorn temperature, which corresponds
to $R=3/2$. We also comment on the behavior in the intermediate region 
$3/2<R<2$.

\subsec{Low temperature thermodynamics, $R>2$}

The canonical partition sum $Z(\beta)$
is obtained by computing the string path integral with Euclidean
time compactified on a circle of circumference $\beta=2\pi R$.
The precise relation is
\eqn\cantherm{\log Z(\beta)\equiv-\beta F(\beta)=Z_{\rm string}(R),}
where $F$ is the free energy and 
$Z_{\rm string}$ is the string partition sum, which is given by a  
sum over connected Riemann surfaces.
For two dimensional string theory, below the Kosterlitz-Thouless
temperature (\ie\ for $R>2$ in our units) one can ignore winding
modes, \ie\ set $t_n=0$ in \conepertt, and the free energy \cantherm\
is given by the familiar formula \FrenO
\eqn\lowtz{Z_{\rm string}(R)=-{R\over2}\mu^2\log{\mu\over\Lambda}
-{1\over24}(R+{1\over R})\log{\mu\over\Lambda}+\cdots.}
We have written the first two terms in the $1/\mu$ (or genus)
expansion. $\Lambda$ is a large scale $\Lambda>>\mu$ corresponding
to a UV cutoff.

Comparing to \cantherm, we see that the free energy $F(\beta)$
is given by
\eqn\frenbet{F(\beta)={1\over2\pi}\left({1\over2}\mu^2
\log{\mu\over\Lambda}+{1\over24}(1+{1\over R^2})\log{\mu\over\Lambda}
+\cdots\right).}
The temperature independent terms in \frenbet\ can be attributed
to vacuum energy. Indeed, the energy
\eqn\enbet{E={\partial(\beta F)\over\partial\beta}=
{1\over2\pi}\left({1\over2}\mu^2
\log{\mu\over\Lambda}+{1\over24}(1-{1\over R^2})\log{\mu\over\Lambda}
+\cdots\right)}
goes as $R\to\infty$ to the ground state energy
\eqn\enotzero{E_0={1\over2\pi}\left({1\over2}\mu^2
\log{\mu\over\Lambda}+{1\over24}\log{\mu\over\Lambda}
+\cdots\right)}
and it is natural to subtract it from both $E$ and $F$.
This leads to
\eqn\efbetcon{\eqalign{E=&-{1\over2\pi}{1\over24 R^2}
\log{\mu\over\Lambda}+\cdots\cr
F=&+{1\over2\pi}{1\over24 R^2}
\log{\mu\over\Lambda}+\cdots.\cr}}
We can now also use the thermodynamic relation
\eqn\SFBE{-\beta F=S-\beta E}
to determine the entropy,
\eqn\sgastwo{S=\beta(E-F)=-{1\over12R}\log{\mu\over\Lambda}+\cdots.}
Eqs. \efbetcon, \sgastwo\ have an obvious interpretation:
they correspond to thermodynamics of a single massless $1+1$
dimensional field which lives in a spatial volume
\eqn\volliouv{V_L=-\log{\mu\over\Lambda}.}
Indeed, solving \efbetcon\ for $R$ in terms of $E$, $V_L$, and
substituting in \sgastwo, one finds
\eqn\sonesc{S=\sqrt{2\pi E V_L\over6}}
which is the standard result for a single massless scalar field
in $1+1$ dimensions. The scalar field in question is of course
the massless ``tachyon'' of $1+1$ dimensional string theory, the
only perturbative field theoretic degree of freedom. Thus, we conclude
that the low temperature thermodynamics corresponds to a gas of
tachyons.

\subsec{Near-Hagedorn thermodynamics, A: $R>3/2$}

As discussed in the introduction, the high energy density 
of states of two dimensional string theory has the Hagedorn
form \rhoeb. Thus, it is interesting to consider the
thermodynamics in the vicinity of the Hagedorn temperature
$T_H=1/\beta_H$. As we show below, $\beta_H=2\pi R_H$,
with $R_H=3/2$. The thermodynamics in the vicinity of the
Hagedorn temperature is quite sensitive to the value of
the constant $s_1$ in \rhoeb. We will see below that 
\eqn\sonerange{-2<s_1<-1.} 
For now we will take \sonerange\ for granted and study 
its consequences for the thermodynamics
at temperatures slightly below $T_H$. In the next subsection
we discuss the situation for temperatures slightly above
$T_H$, and in the process determine both $T_H$ and $s_1$.

Consider the canonical partition sum
\eqn\HAGED{ Z(\beta)=\int_0^\infty \ dM \rho(M) \ e^{-\beta M}.}
The high energy behavior of $\rho(M)$, given by \rhoeb, implies
that the integral converges for $T<T_H$, but the exponential
Boltzmann suppression disappears as $\beta\to\beta_H$. Naively, 
one might expect that in this limit the integral \HAGED\ becomes
dominated by high energy states, so that
\eqn\zzzbbb{Z(\beta)\sim (\beta-\beta_H)^{-s_1-1},}
but this is not quite the case. The contribution of 
the high energy part of the spectrum to $Z$ is given in \zzzbbb. 
The rest of the states give an additive contribution $Z_{\rm low}(\beta)$ 
which is analytic as $\beta\to\beta_H$ and approaches a non-zero
constant in this limit. The partition sum \HAGED\ can
thus be written for $\beta$ slightly larger than $\beta_H$ as 
\eqn\zzssww{Z(\beta)=Z_{\rm low}(\beta)+c(\beta-\beta_H)^{-s_1-1},}
where $c$ is a constant. Since $s_1+1$ is negative (see \sonerange), 
the contribution of the high energy states (the last term in \zzssww) 
goes to zero, and the partition sum \HAGED\ is in fact dominated by 
states with moderate energies. 

Naively, the Hagedorn temperature appears to be a limiting
temperature, since the energy as measured in
the canonical ensemble diverges:
\eqn\mmeeaa{E=-{\partial\log Z\over\partial\beta}\sim
(\beta-\beta_H)^{-s_1-2}.}
Note that $-s_1-2$ is negative (see \sonerange); thus the derivative
of the second term on the r.h.s. of \zzssww\ is much larger
than that of the first. Hence, it appears that one needs to supply 
an infinite amount of energy to heat up the system to the Hagedorn 
temperature.

However, one has to be careful with this conclusion, since
the whole notion of the canonical ensemble breaks down as
one approaches the Hagedorn temperature from below, since
the fluctuations around the mean energy are large:
\eqn\enrgfluc{
{\langle E^2\rangle-\langle E\rangle^2\over\langle E\rangle^2} 
\sim (\beta-\beta_H)^{s_1+1}.}
In such situations the system should be studied in
the microcanonical ensemble, where the energy is fixed,
and the temperature is given by
\eqn\microtemp{\beta={\partial S\over\partial M}.}
Using the entropy \rhoeb\ we find 
\eqn\microbeta{\beta=\beta_H+{s_1\over M}+O\left({1\over M^2}\right).}
Since $s_1$ is negative (see \sonerange), we find that
at high energies the temperature is {\it above} the Hagedorn temperature.
Clearly, this does not mean that temperatures below the Hagedorn
one are not achievable in two dimensional string theory -- as we
reviewed in the previous subsection, the low temperature
thermodynamics is perfectly sensible. Instead, \microbeta\ implies 
that the microcanonical thermodynamics slightly below $T_H$ corresponds 
to a finite energy $M$, obtained by solving \microtemp. 

Thus, if we imagine a process where two dimensional string theory
is gradually heated up by supplying energy to the system in the
microcanonical ensemble, the temperature $T_H$ is reached at a
finite value of the energy $M$.
This suggests that the Hagedorn temperature is associated with
a phase transition (as opposed to being a limiting temperature). 

\subsec{Near-Hagedorn thermodynamics, B: $R<3/2$}

In the previous subsection we saw that just below
the Hagedorn temperature the thermodynamics is dominated by states 
with moderate energies. Since two dimensional black holes should only
dominate the thermodynamics at large energies, it is clear that
the low temperature phase is not well described by perturbative
string theory in the Euclidean black hole background \aaa, even
as $T\to T_H$. Instead, as we show in this subsection, string 
theory in the background \aaa\ describes the thermodynamics just 
above the Hagedorn temperature.

For very large $M$, the thermodynamics is obtained by
studying {\it classical} string theory on the cigar. The
mass $M$ is related to the value of the string coupling
at the tip of the cigar, while the temperature is determined
by the asymptotic radius of the cigar. Since the latter is 
independent of the former, we conclude \GP\ that
\eqn\clasbet{\beta={\p S\over\p M}=\beta_H=2\pi{3\over2}}
is constant and the entropy-energy relation is
\eqn\entenclass{S=\beta_HM}
\ie\ the density of states exhibits Hagedorn growth,
$\rho(M)\sim \exp(\beta_HM)$ at large $M$. The free energy 
\SFBE\ should thus vanish in the classical approximation.

\lref\GibHaw{G. W. Gibbons and S. W. Hawking,
Phys. Rev. {\bf D15} (1977) 2752.}

Quantum string theory on the cigar leads to $1/M$ corrections to \clasbet,
\entenclass, whose general form is (see \rhoeb)
\eqn\ENTRR{\eqalign{
S=& \beta_H M+s_1\log M+O\left({1\over M}\right)\cr
\beta=&\beta_H+{\alpha\over M}+O\left({1\over M^2}\right),\cr
}}
where $s_1$, $\alpha$ are constants to be determined.

For $\beta<\beta_H$, the integral \HAGED\ defining the
canonical partition sum diverges. At first sight this 
looks like a problem, but in fact it is standard in
black hole theory. For example, black holes in flat spacetime
generally have an entropy that grows like $M^a$ with
$a>1$, so the analog of \HAGED\ diverges for all temperatures.
The Gibbons-Hawking prescription \GibHaw, whose generalization
to string theory we are studying here, defines the canonical
partition sum by a formal saddle point evaluation of the
integral \HAGED. The imaginary part of the resulting
partition sum is associated with the thermodynamic instability
of these black holes (their negative specific heat), and is
related to the appearance of an unstable mode (a tachyon)
in the Euclidean black hole background. 

We will see that something very similar happens here. Two dimensional
string theory is thermodynamically unstable for $T>T_H$. Nevertheless,
just like in \GibHaw, we can compute the free energy by a formal saddle
point evaluation of \HAGED. The Euclidean black hole background has
a tachyon, whose condensation needs to be understood in order to
study the high temperature phase of two dimensional string theory.

It is not difficult to show that a saddle point evaluation of the
integral \HAGED\ with the density of states \rhoeb\ gives the result
\zzzbbb. The saddle point is located at $M$ given by \microbeta,
and assuming that $s_1$ is indeed negative \sonerange, as we will
find soon, we see that for $\beta$ slightly below $\beta_H$ it is
located at a large positive value of $M$, which is consistent with
our use of the asymptotic density of states \rhoeb\ to find it.

Since the partition sum is given by \zzzbbb, the free energy is
\eqn\fffeee{-\beta F=-(s_1+1)\log(\beta-\beta_H)+O(1).}
The energy-temperature relation (in the canonical ensemble)
is obtained from \fffeee\ by using
\eqn\meanE{M={\p(\beta F)\over\p\beta}={s_1+1 \over\beta-\beta_H}.}
Note that it is different from the relation \microbeta, which 
is valid in the microcanonical ensemble. The two are not the
same here because of the large energy fluctuations in the canonical
ensemble.

Substituting \meanE\  in \fffeee\ we conclude that
\eqn\ffbbnn{-\beta F=(s_1+1)\log M+O(1).}
Thus, by computing the string partition sum \cantherm\ we can deduce the
entropy-energy relation characterizing the black hole, \ENTRR. 

An immediate puzzle is that while we expect the tree level
contribution to the free energy (the term that goes like $M$ in
\ffbbnn) to vanish, our explicit calculation in section 5 gave in fact
a finite answer for this coefficient. This apparent paradox is
resolved by noting \GP\ that the thermodynamic relation \SFBE\ is
modified in this case by the presence of a massless scalar field, the
dilaton, that mediates a long range force. The black hole spacetime is
in fact charged under the dilaton current $J_a=\epsilon_a^b\nabla_b
e^\phi$, and therefore the thermodynamic relation \SFBE\ contains a
correction that is proportional to the dilaton charge of spacetime. It
was shown in \GP\ that this gives rise to a term in the free energy
that goes like $M$ (times a constant that is ambiguous in the gravity
approximation), and
should be interpreted as a chemical potential for the dilaton charge.

Therefore, we believe that the correct procedure for studying the
thermodynamics of two dimensional string theory is to subtract this
term from $F$. Note that in the Sine-Liouvile description, this
is a term that goes like $\lambda^8$ (see \BHLIM). In general in
non-critical string theory, such terms are due to physics far from the
Liouville (or in this case Sine-Liouville) wall and are ``non-universal''
(as noted at the end of section {\it 5.1}). 
In our particular case one could alter the coefficient of this term by
adding a constant (independent of $\mu$) to the eight point function
$\langle \CT_R^4\CT_{-R}^4\rangle$ given in
\MOOR; such a constant is non-universal in Liouville theory (or
the double scaled matrix model).  

As further justification for the validity of this
procedure we discuss in the next section a generalization 
of two dimensional string theory to the fermionic case (with worldsheet
supersymmetry but {\it no} spacetime supersymmetry). The physics
of this theory is very similar to that of the bosonic one, but the
tree level partition sum vanishes.  

The one loop correction to the free energy discussed in section {\it 5.3}
(see \TORLIM) implies that 
\eqn\sssone{s_1+1=-(R+R^{-1})/48=-13/288.}
which is in agreement with \sonerange.

The high temperature thermodynamics described by the black hole is 
unstable. As is clear from \meanE, \microbeta, the specific heat is 
negative: as the energy increases, the temperature decreases towards 
$T_H$. This is of course closely related to the fact that in the
formal saddle point evaluation of the partition sum
\HAGED, we are in fact expanding around a maximum of
$f(M)=\beta M-S(M)$ (rather than a minimum). This
thermodynamic instability is associated with
the presence of a negative mode in the Euclidean 
path integral, \ie\ one expects to find a {\it tachyon}
in the cigar background \aaa. 

Because the instability of the thermodynamics
is a one loop effect in the cigar background, one actually
expects to find a zero mode in the classical theory,
which becomes a negative mode after one loop effects
are included. There indeed is a very natural candidate
for precisely such a mode which we will describe next.
This issue was recently discussed in the context of LST
in \kkss, so we will be brief. 

It is known (see \eg\ \GK) that in the cigar geometry 
there are normalizable states localized near the tip,
corresponding to the principal discrete series representations
of $SL(2)$, \ie\ for
\eqn\prindis{|m|=j+l;\;\;l=1,2,3,\cdots.}
One way to find these states is to study the correlation functions
of the non-normalizable observables $V_{j;m,\bar m}$ (see section
2); the states \prindis\ give rise to poles in these correlation
functions. The classical zero mode (or ``massless state'')
referred to above corresponds to\foot{Since the result is
general, we are writing it as a function of $k$. For our case,
as usual, $k=9/4$.}
\eqn\mjcomb{m=\bar m={k\over2};\;\;j={k\over2}-2.}
It is easy to check that this combination of $j$ and $m$ satisfies
$\Delta_{j;m,\bar m}=\bar\Delta_{j;m,\bar m}=1$ (see \scdimv), 
\ie\ this is an
on shell masless state (a zero mode). As discussed in \kkss, it is 
likely that including one loop effects turns it into a negative mode.
It would be interesting to prove this by an explicit one loop
calculation. This should be possible using the results of section
5. 

Thus, we conclude that our perturbative analysis of the black hole
background describes the high energy phase of two dimensional
string theory. The thermodynamics is unstable, and there is 
a negative mode (or more precisely a classical zero mode that
is conjectured to become unstable when one loop effects are taken
into account). The high temperature phase presumably involves 
condensation of this mode and would be interesting to study
using the techniques of sections 4, 5.

\subsec{The intermediate region $3/2<R<2$}

So far we have discussed the thermodynamics
below the Kosterlitz-Thouless temperature, \ie\ for $R>R_{KT}=2$,
where it is dominated by a gas of perturbative
string states (see \efbetcon\ -- \sonesc), and at temperatures
near the Hagedorn temperature. This leaves an unexplored 
intermediate region, $3/2<R<2$. We will not study this region 
in detail here, but would like to make a few comments on it.

As the temperature is raised above the Kosterlitz-Thouless
one, the system is expected to undergo a phase transition to a
phase where the non-singlet degrees of freedom contribute to the
thermodynamics \GRKL. In this phase, the free energy and entropy
are expected to receive tree level contributions (in contrast to
\efbetcon, \sgastwo), associated with the large density of 
non-singlet states.

\lref\seishen{N. Seiberg and S. Shenker, hep-th/9201017,
Phys. Rev. {\bf D45} (1992) 4581.}

In the continuum description, one expects the winding modes \conepertt\
to condense. For $1<R<2$, only the winding number one mode is
relevant, and thus we expect $t_{\pm1}=\lambda/2$ in \conepertt\
to be the only non-zero couplings, in addition to $\mu$. An important
question for analyzing the thermodynamics is how does $\lambda$
depend on $R$. It is not clear to us how one can determine\foot{Naively,
$\lambda$, $R$ are two independent parameters that specify superselection
sectors in the theory. However, as we saw above and discuss further 
in section 8, it is important here to keep the UV cutoff $\Lambda$ finite.
This makes $\lambda$, $R$ fluctuating fields, and one must extremize
with respect to them. See \seishen\ for a related discussion.}
$\lambda(R)$, but it appears likely that it is such that the
parameter $z$ defined in \defzz\ varies smoothly between zero and 
some finite value, 
as $R$ decreases from $R=2$ to $R=3/2$ (the Hagedorn temperature). 
This variation is related to the dependence of the energy
on the temperature. As $R\to 3/2$ the energy should approach
a finite value (as discussed above), while as $R\to 2$ we 
should connect to the low temperature thermodynamics, for which $z=0$. 

The qualitative picture proposed
in the previous paragraph is in agreement with the
phase diagram of the Sine-Gordon model (or, in statistical mechanics
language, the Kosterlitz-Thouless phase diagram). 
In the region $R<2$ the phase diagram is such that as the distance
scale increases (\ie\ as one goes from the UV to the IR), $\lambda$
grows and $R$ decreases. String thermodynamics for $3/2<R<2$ can be
thought of as the Sine-Gordon model coupled to worldsheet gravity. 
It is well
known that the qualitative structure of the RG in two dimensional QFT 
coupled to gravity is the same as in the theory without gravity, with
$\mu$ playing the role of the RG (energy) scale. $\mu\to\infty$ corresponds
to the UV, while small $\mu$ is the IR region. Thus, in the Sine-Gordon model
coupled to gravity one expects to find $\lambda=\lambda(\mu)$ and $R=R(\mu)$,
\ie\ $\lambda=\lambda(R)$. It is not difficult to see that the qualitative
properties obtained from this picture are in agreement with the proposal in
the previous paragraph. A more precise analysis is left for future work.

\newsec{Fermionic two dimensional string theory}

In this section we briefly describe a natural generalization
of two dimensional string theory to the fermionic (type 0) case.
Our main motivations for discussing this theory are the following.

In section 6 we argued that in the bosonic $2d$ string, the tree
level free energy that goes like the mass of the black hole, corresponds
to a vacuum energy type contribution, and should be subtracted. We will
see that most of the physical properties of the type 0 theory are
very similar to the bosonic one, but the tree level free energy
vanishes. This provides further evidence for the validity of the
procedure of section 6.

The second reason for studying this generalization is that it
provides a bridge to higher dimensional physics. In particular,
the two dimensional worldsheet supersymmetric black hole
plays an important role in the study of the thermodynamics of
Little String Theory (see \kkss\ for a recent discussion). 

Two dimensional type 0 string theory contains two worldsheet
superfields, $X$, with central charge $c={3\over2}$ and $\phi$, 
with $c={3\over2}+ 6Q^2$. The fact that the total central charge 
is $c=15$ determines $Q=\sqrt2$. 
The zero temperature physics of the theory is described in \difkut.
The perturbative spectrum includes in this case two massless scalars, 
a NS-NS field $T(X)$, and a R-R field $V(X)$. As in the bosonic case, 
there are no perturbative massive string states (for
generic momenta). The bulk amplitudes 
of $T$ and $V$  are known; they are completely determined by an infinite
dimensional symmetry algebra, which is very similar to that found in the
bosonic string, and are themselves very similar to the bosonic ones. 

Our main interest here is in the high temperature thermodynamics
of the model. Just like in the bosonic case, it should be dominated
by states associated with $1+1$ dimensional black holes. The black 
hole background coresponds in this case to an $N=1$ superconformal
version of the coset \cosetconf. Algebraically, it can be thought of
as follows.

One starts with the $N=1$ superconformal theory on $H_3^+$, which contains 
an $SL(2,R)_L\times SL(2,R)_R$ current algebra with total level $k$. The
level $k$ receives a contribution of $k+2$ from a bosonic WZW on $H_3^+$,
and $-2$ from three free (left and right moving) fermions in the adjoint 
of $SL(2)$. The total central charge of the theory is
\eqn\cfermH{c_H={3(k+2)\over k}+{3\over2}.}
One then gauges a $U(1)$ super-Kac-Moody algebra, thereby decreasing
the central charge to 
\eqn\ccoset{c_{\rm coset}={3(k+2)\over k}.}
The requirement that the total central charge is equal to $15$ determines
\eqn\kQrel{k={1\over Q^2}={1\over2}.}
Much of the discussion of section 6 goes through here. The classical
analysis gives\foot{We work in the same units as in the bosonic theory.
The Kosterlitz-Thouless radius is $R=\sqrt2$ in the fermionic case.} 
\eqn\entferm{S=\beta_H M;\;\;\beta_H=2\pi\cdot{1\over\sqrt2}.}
Therefore, the classical free energy \SFBE\ should vanish. Unlike
the bosonic case, here this can indeed be shown to be the case,
as we briefly review next (see \kkss\ for further discussion). 

As discussed in section 5, in the bosonic Euclidean cigar background, 
the spherical partition sum (which is related to the classical  
free energy via \cantherm) is in general non-zero. The volume of the cigar 
cancels the volume of the Conformal Killing Group of the sphere and leaves
behind a finite residue. 

In the superstring one has to take into account fermionic zero modes.
The CKG is replaced in this case by a supergroup, SCKG, but the associated
fermionic zero modes are cancelled by fermionic zero modes of the $N=1$
SCFT on the cigar. Thus, it appears that in the fermionic string, the
partition sum in the background \cosetconf\ is non-zero as well. 

This conclusion is incorrect because of the fact, mentioned at the end
of section 2, that the SCFT on the cigar has in fact a larger symmetry,
an accidental $N=2$ superconformal symmetry. Thus, this SCFT has twice
as many fermionic zero modes as one would guess, and integrating over them
leads to a vanishing spherical partition sum \kutsei. 

The vanishing of the free energy in this case is due to $N=2$ worldsheet
supersymmetry. It does not require spacetime supersymmetry, which in
fact is absent in the type $0$ theory under consideration. It is also
interesting to note that the free energy only vanishes at the Hagedorn
temperature. This can be seen by using the duality of \GK\ relating the
Euclidean cigar SCFT to $N=2$ Liouville theory. As discussed in \GK\ for
$k<1$, which is the case here \kQrel, the $N=2$ Liouville is more
weakly coupled and thus more appropriate. 

As one changes the radius of Euclidean time, the $N=2$ Liouville
superpotential changes to one that only preserves an $N=1$ subalgebra
of the $N=2$ superconformal symmetry. Thus, the arguments above no
longer imply that the free energy vanishes, and one expects 
that it does not. This is in agreement with the thermodynamic picture
that one expects: the free energy goes to zero as the energy goes to 
infinity, or $\beta\to\beta_H$. At finite energy the free
energy is non-zero.

One can also repeat for this case the calculation of the one loop
partition sum in section {\it 5.4}, and the discussion of section 6,
with very similar conclusions. To go beyond one loop probably
requires finding a matrix model that describes the dynamics. This
is an interesting open problem. 
 
\newsec{Some open problems and future directions}

In this paper we discussed 
the thermodynamics of two dimensional string theory above the
temperature corresponding to the Kosterlitz-Thouless phase 
transition. In the Euclidean time formulation of the problem, 
this regime is characterized by the condensation of winding modes
around Euclidean time. We showed that the theory undergoes a 
finite temperature phase transition around the Hagedorn
temperature $T_H$. The high temperature phase near $T\simeq T_H$
is well described by the Euclidean two dimensional black hole. 
Furthermore, by using the conjectured correspondence \FZZ\ of 
string theory in the black hole and Sine-Liouville backgrounds, 
we constructed a matrix model which provides a powerful tool for
studying the two dimensional black hole to all orders in string 
perturbation theory.

There are many open problems associated with our work,
some of which are listed below.

While the differential equation \TodaD\ in principle
determines the partition sum $\CF(\lambda,\mu)$ to all orders in
string perturbation theory (as described in section 5), we have 
so far not been able to solve the resulting equations \ANYH\ beyond 
genus one. Furthermore, \TodaD\ is just the lowest equation in an 
infinite hierarchy which should allow the 
calculation of correlators of winding modes with arbitrary winding 
number at any order in $g_s$. Some low genus correlation functions, 
like \twoptvjm, have been computed in the continuum formalism, and 
it would be nice to reproduce them using the Hirota equations (see 
appendix C). These calculations remain an interesting open problem.  

There are observables in two dimensional string theory which seem to be
outside of the framework  of the Toda equations. Examples include modes
carrying momentum around Euclidean time, and discrete states. It would be
interesting to generalize the formalism to include these couplings.

In the discussion of black hole thermodynamics in section 6,
we assumed that the mass of the black hole is related to the string
coupling at the tip of the cigar via the classical relation $M\sim
\exp(-2\Phi_0)$. This relation is expected to be corrected quantum
mechanically, and such corrections will influence the thermodynamic
relations at two loop level and beyond. Thus, this issue will need
to be addressed when the higher order results mentioned above
become available.

It would be interesting to understand better the thermodynamics of two
dimensional string theory in the intermediate region $3/2<R<2$, as well
as in the high temperature phase $R<3/2$, and in
particular to explore further the relation to the Kosterlitz-Thouless
phase diagram mentioned in section 6.

The FZZ correspondence \FZZ\ described in section 2 played an
important role in our discussion, since it allowed us to replace the
curved geometry of the cigar, which is difficult to realize in MQM, by
the winding mode condensate which is easier to study in the matrix
model. This correspondence remains very mysterious. It is not clear
how the metric of the cigar is encoded in the Sine-Liouville
potential. The phenomenon is probably much more general (\eg, as we
mentioned, it can be extended to the fermionic string) and should be
very interesting to understand better.

String thermodynamics seems to suggest that two dimensional string
theory\foot{Note that here we mean the standard $c=1$ string theory at
zero temperature and with the scale set by the cosmological constant
$\mu$.} has a Hagedorn spectrum of states at high energies, given by
eq. \rhoeb. This is surprising, since the perturbative spectrum
consists of one field theoretic degree of freedom. Thus, it must be
that most of the states described by
\rhoeb\ are non-perturbative, and it would be very interesting
to understand what they are.\foot{We note that a very similar situation
occurs in weakly coupled Little String Theory in the double scaling
limit defined in \GK. The perturbative spectrum has a Hagedorn density
of states of the form \rhoeb, but thermodynamics suggests that the
full theory has a richer spectrum. In particular, $\beta_H$ is
predicted to be larger than the perturbative one. Thus, most of the
states are non-perturbative.} A natural proposal is that these
states belong to the non-singlet sector of the matrix model. The
degeneracy of non-singlet states is very large, but their masses are
proportional to $\log(\Lambda/\mu)$ (with $\Lambda$ a UV cutoff)
\GRKL\ and thus diverge in the limit $\Lambda\to\infty$.  It appears
that one can define two dimensional string theory  in two different
ways.

If one sends $\Lambda$ to infinity, all the non-singlet states are
removed from the spectrum. The matrix dynamics reduces to that of just
the singlets (\ie\ the eigenvalues of $M$ \GENF), which in spacetime
means that the massless ``tachyon'' is the only physical degree of
freedom. The low temperature thermodynamics reviewed in the beginning
of section 6 applies then for all temperatures, and all the vortex
couplings $t_n$ in \conepertt\ are set to zero.  We believe that the
resulting theory does not admit two dimensional black holes \aaa\ and
does not have a Hagedorn density of states.

A second possibility is to keep the UV cutoff $\Lambda$ large but
fixed.  Then the energy gap to the non-singlet states remains finite,
and the physics is much richer. In particular, the model does contain
black holes, and the corresponding Hagedorn spectrum of states
\rhoeb\ should be reproduced by counting non-singlet states in MQM.

\lref\adscft{J. Maldacena, hep-th/9711200, Adv. Theor. Math. Phys. 
{\bf 2} (1998) 231.}

It may seem puzzling that there is more than one way to define two
dimensional string theory, but in fact this situation is standard in
holographic theories that were studied in recent years. Consider, for
example,\foot{Similar comments can be made for $d>3$, but the case
$d=3$ is special in that three dimensional gravity without matter
exists as a quantum theory, and that the theory has an infinite
dimensional symmetry, like two dimensional string theory.} string
theory on $AdS_3$. As is well known, three dimensional gravity with
negative cosmological constant can be quantized as a Chern-Simons
theory. According to the AdS-CFT correspondence \adscft, this theory
is equivalent to a two dimensional CFT with a large central charge (in
the semiclassical limit), which in this particular case contains only
the conformal block of the identity. The high energy density of states
of this theory is much smaller than that given by (BTZ) black hole
thermodynamics; thus this theory does not contain $2+1$ dimensional
black holes.

An alternative definition of gravity on $AdS_3$ is obtained by
studying string theory on $AdS_3$ and including all the perturbative
and non-perturbative string excitations. This gives rise to a much
richer theory, and in particular one can show that it does have black
hole states and their entropy can be computed microscopically, by
studying the CFT dual to string theory on $AdS_3$. This CFT can be
formulated (roughly) as the extreme infrared limit of a matrix field
theory; the bulk of the black hole entropy comes from off-diagonal
components of the matrices. Moreover, just like in our case, the mass
gap to the lowest lying black holes is non-perturbative in the string
coupling.

In two dimensional string theory, the above analogy suggests that we
should keep $\Lambda$ large but fixed in the double scaling limit.
This is in fact very natural since $\Lambda\sim N$, so we are
proposing that the non-perturbative theory has large but finite
$N$. This is again similar to string theory on AdS space, where the
analogs of $N$, $\mu$ are kept large but fixed.\foot{These analogs are
the radius of curvature of the AdS space in string units, and in
Planck units.}

\lref\KAZMA{ V. A. Kazakov, Solvable Matrix Models, proceedings of the 
MSRI Workshop ``Matrix Models and Painlev\'e Equations'', Berkeley
(USA) 1999; hep-th/0003064.}

The large entropy \rhoeb\ should then be due to non-singlet
states. Not much is known about the spectra of such states at present, 
especially in the large $N$ limit of interest here. One approach to 
this problem is to rewrite the partition function \AVTW\ as a sum of 
Gibbs partition functions over representations $r$ of $SU(N)$:
\eqn\SUMR{
{\cal Z}_N(\beta , \lambda ) = \sum_r g_r[\lambda] Tr_r e^{-\beta
{\widehat H}_r},}
where 
\eqn\GRAT{g_r[\lambda]=
 \int [D\Omega]\chi_r(\Omega^\dagger)
 \exp\left( \sum_{n\in\Z} \lambda_n \tr
 \Omega^n\right), }
($\chi_r(\Omega^\dagger)$ is the Weyl character in the representation 
$r$) and
\eqn\HRP{
{\widehat H}_r = P_r \sum^N_{k = 1} \big[ -{1 \over 2}
{\partial ^2 \over \partial x_k^2} - {1 \over 2} x^2_k \big]
+ {1 \over 2} \sum_{i \neq j}
{{\widehat \tau }^r_{ij}{\widehat \tau }^r_{ij} \over (x_i - x_j)^2}
}
is the Hamiltonian in representation $r$ in terms of the eigenvalues $x_i$
of the matrix $M$ with the inverted oscillator potential. This
Hamiltonian is a matrix in the representation space: $P_r$ is a
projector to this space and ${\widehat \tau }^r_{ij}$ are $SU(N)$
generators on $r$ (see \refs{\BULKA,\KAZMA} for details). These
Hamiltonians define, in principle, in a natural way the time evolution
of the system. Therefore to count the states in a given energy interval 
one can try to diagonalize them for representations
corresponding to large Young tableaux. This appears to be difficult
due to the Calogero type interaction of the eigenvalues with an $SU(N)$
spin structure (see \refs{\GRKLbis,\BULKA} for a discussion of the
adjoint representation). 

If the non-singlet states indeed account for the entropy of black hole
\rhoeb, it is natural to ask whether one can think of non-singlets
as excitations of black holes directly. A possible direction for
making the relation more precise is to note that the non-singlet
states can roughly be thought of as open strings. In black hole
physics it has been proposed in the past that the black hole entropy
is due to open strings whose ends lie on the horizon of the black
hole. It would be interesting to make this relation more precise.

An interesting set of issues is associated with the Minkowski
continuation of our results. The description of the black hole in
terms of the Sine-Liouville model involves winding modes around
Euclidean time and thus seems inherently Euclidean.  Yet, many of the
interesting questions about black hole physics can only be discussed
in the Minkowski version. Examples are the unitarity of scattering off
black holes and the understanding of the dynamics of black hole
formation. In principle, the Hamilonians \HRP\ provide a direct way
for the analytical continuation to Minkowski space.  Thus, to understand
the formation and Minkowski evolution of black holes one needs
a better understanding of the dynamics of the non-singlet states. 
This will have to be left for future work. However, we would like 
to comment on one point related to black hole formation in two 
dimensional string theory.

It is sometimes stated in the old matrix model literature that the
black holes \aaa\ are not good toy models of higher dimensional black
holes, since they cannot be formed in two dimensional string theory
because of the large symmetry present in this theory.\foot{For a review
of gravitational physics in two dimensional string theory see
\POLCHINSKI.} This $W_\infty$ symmetry (see
\refs{\KLEBANOV,\ginsmoore}) is closely related to the interpretation
of tachyon dynamics in terms of free fermions, with the tachyon
describing ripples on the Fermi surface. The fact that the underlying
fermions are free does not allow black holes to form in the evolution
of tachyon pulses, even if they are large.

Our discussion in this section clarifies the extent to which this
statement is expected to be correct.  In the theory with
$\Lambda\to\infty$, which corresponds to free fermions and describes
the dynamics of tachyons as perturbations of the Fermi surface, one
indeed does not expect to be able to make black holes since the theory
does not contain black holes. However, in the theory with finite
$\Lambda$, the free fermions are just one sector of the Hilbert space
(corresponding to singlet dynamics) and black hole formation is
expected to occur in the collapse of matter made out of non-singlet
states.

Note that we are not proposing that the $W_\infty$ symmetry of
perturbative two dimensional string theory is broken
non-perturbatively.  Rather, we are saying that it does not prevent
the appearance of non-trivial dynamics in the non-singlet sectors. As
mentioned above, the situation is analogous to string theory on
$AdS_3$. There, the infinite dimensional spacetime Virasoro algebra
(an analog of the $W_\infty$ symmetry of two dimensional string
theory) is a symmetry of the full theory, but it does not prevent
non-trivial dynamics and black hole formation in general, although the
conformal block of the identity indeed does not contain black hole
states (just like the singlet sector in two dimensional string
theory).

\lref\KS{D. Kutasov and N. Seiberg, Nucl. Phys. {\bf B358} (1991) 600.}
\lref\niarchos{V. Niarchos, hep-th/0010154.}

Finally, we would like to mention another aspect of our results,
related to the Hagedorn spectrum \rhoeb. In perturbative string
theory, a spectrum of this sort would lead to an infrared instability
(a tachyon), unless there was an almost complete cancellation between
bosons and fermions \refs{\KS,\niarchos}. It is an interesting open
question whether this relation between the density of states and IR
stability is more general. Assuming that it is, it would be interesting
to understand how to reconcile it with the behavior found here.

One possibility is that the theory actually has fermionic excitations.
Perturbative two dimensional string theory has no spacetime fermions;
however, the statistics of the non-perturbative states is not obviously 
bosonic. Recall that in the singlet sector, the underlying degrees of 
freedom are free fermions. Another possibility is that the actual high 
energy density of states is in fact much smaller than that suggested by 
\rhoeb. Further discussion of this possibility
requires a better understanding of the high temperature phase of the
theory (associated with tachyon condensation - see section 6). These
issues deserve a better understanding.

\bigskip
\noindent{\bf Acknowledgements:}
We thank S. Alexandrov, D. Gross, E. Martinec, A. Polyakov,
D. Sahakyan, S. Shenker and L.  Susskind for useful discussions. 
V.K. thanks A. Polyakov for attracting his attention to 
 the conjecture of \FZZ\  and  to the  possibility opened by 
 this coinjecture to describe  the black hole within the compactified 
 c=1  matrix model. The
research of I.K and V.K. is supported in part by European network
EUROGRID HPRN-CT-1999-00161. The work of D.K. is supported in part by
DOE grant \#DE-FG02-90ER40560. V.K. thanks the University of Chicago and
Stanford University for hospitality during the course of this
work. D.K. thanks the ITP at Stanford University for hospitality in
the final stages of this work.  

\appendix{A}{Free compactified fermions from the MQM  
averaged over twist variable}

To show how free fermions emerge from the MQM partiton function
\GENF\ let us  use the following formula:
\eqn\KMM{ \CZ_N^{singlet}= 
\int [d \Omega]_{SU(N)} \CZ_N(\Omega)= \int \CD M \int 
[d\Omega]_{SU(N)} K_{2\pi R}(M,\Omega^+M\Omega),}
where $K_{2\pi R}(M,M')$ is the propagation kernel for the matrix
variable from the initial value $M$ to the final value $M'$ during the
time $2\pi R$.
    
Using the discrete time version of \GENF\ with the action given by
\eqn\DISCF{  S=\sum_k \Tr\left( {1\over 2 a} 
(M_k-M_{k-1})^2 + a V(M_k)\right),}
diagonalizing the matrices and applying consecutivly the
Itzykson-Zuber formula to integrate out the angular variables \KAZREV\
we obtain in the limit of a small lattice spacing $a\to 0$ the
following representation for the propagation kernel of the eigenvalues
$u_1,\cdots,u_N$:
\eqn\KUU{ K_{2\pi R}(\{u\},\{u'\}) =  {1\over \Delta(u)\Delta(u')}
{det}_{kj}K_{2\pi R}(u_k,u'_j) }
where $\Delta(u)=\prod_{k>j}(u_k-u_j)$ is the Vandermonde
determinant.  We ignored in \KUU\ the overal normalization constant.
Finally putting $M=M'$ (which means $u_k=u'_k$) and integrating with
respect to the eigenvales of the final matrix $M$ with the Dyson
measure $\Delta^2(u)\prod_{k=1}^N \ du_k$ we obtain from \KMM\ and
\KUU\ the following representation for the singlet partition function:
\eqn\FRFE{  \CZ^{(singlet)}=
\int \prod_{i=1}^N \ d u_i \ \ \ {det}_{kj}K_{2\pi R}(u_k,u_j),   }
or, returning to the functional integral,
\eqn\GENFU{ \CZ_N^{(singlet)}
= \sum_\CP (-1)^\CP \int_{u(2\pi R)=\CP^{-1} u(0)\CP} \ \ \ 
\prod_{k=1}^N{\CD}u(x) e^{ -
\int ^{2\pi R} _0 dx \left[ \hf (\p_x u_k)^2  +V(u_k)\right] }}
where the sum goes over permutations $\CP$ of the eigenvalues
$(u_1,\cdots,u_N)$ in the process of propagation around the circle.

The last formula gives the explicite free fermion representation of
the singlet sector of the compactified MQM.

\appendix{B}{Evaluation of the twisted partition function in the scaling limit}

If we forget for a moment about the wall, the problem decomposes into
$N^2$ independent inverted oscillators.  They are related only by
common twist angles.  For each complex matrix element $M_{kj}$, the
Schr\"odinger equation on the wave function $\Psi(M_{kj})$ looks as
\eqn\SCRD{  -\hf \left({\p^2\over \p M_{kj} \p M_{kj}^*} +  M_{kj}
M_{kj}^*\right)\Psi(M_{kj}) =E\Psi(M_{kj})  }
where there are no summations over the repited indices.  We have to
calculate from here the density of states in the presence of twists.
In the lagrangean formulation we have the periodic boundary conditions
$M_{kj}(2\pi R)=e^{i\t_{kj}} M_{kj}(0)$, where $\t_{kj}=\t_k-\t_j$. To
express them in the hamiltonian language we have to go to the polar
coordinates $M_{kj}=r_{kj} e^{i\t_{kj}}$, where we fixed the angular
dependence by the difference of fixed twist angles, and look for the
wave function in the form
\eqn\ANZPS{ \Psi(r_{kj},\t_{kj})=
\sum_{m=-\infty}^\infty \chi_m(r_{kj})e^{im\t_{kj}}.  }
The density of energy will be given in terms of the sum of densities
$\rho_m(E)$ of the states with a given angular momentum $m$ weighted
by $e^{im\t_{kj}}= e^{im(\t_k-\t_j)}$ (which can be calculated from
the analysis of the quasicalssical wave function, see
\BULKA\ for the details):
$$
\rho(\t, E)=\sum_{m=-\infty}^\infty e^{im\t}\rho_m(E) .
$$
The explicit calculation gives 
\eqn\RHOTW{  \rho(\t, E)={\sinh[(\pi-\t) E]\over \sinh[\pi E]
\sin\t}+\delta(\t)\log\Lambda^2 .  }
Note that this result does not depend on the familiar cutoff of the
$c=1$ string theory for the nonequal twist angles
$\t_{kj}=\t_k-\t_j\neq 0$. The system becomes stabilized due to the
twisting!

Now we can  calculate the partition function of the individual matrix
element in the fixed temperature $T=1/2\pi R$:
\eqn\INDP{   \CZ_1(\t_{kj},R)=\int_{-\infty}^\infty d E
e^{-2\pi R E}\rho(\t_{kj}, E)={1/2\over \cos (2\pi R) - \cos\t_{kj} }.  }
Strictly speaking, this formula  makes sense only for $R<1/2$ and
the distribution of $\theta$'s in the interval
$|\theta_{max}-\theta_{min}|<\pi R$. 
 These limitations are  inherited from the
canonical (fixed $N$) ensemble used so far. We will see that it
is not present in the grand canonical ensemble.

{}From this formula, by taking the product of partition functions of
the individual matrix elements, we derive the partition function of
the twisted inverted oscillator
\eqn\TWPA{   Z_N(\t_1, \cdots , \t_N ,R)=
\left(2\sin{\pi R} \right)^{-N}\prod_{i>j}{1\over \cos (2\pi R) -
\cos(\t_i-\t_j) }    }
where the first factor is the $\t$- independent contribution of the
diagonal matrix elements.

\appendix{C}{  $\tau$-function and Hirota equations for the Toda 
chain hierarchy}

\subsec{Vertex operator representation}

The simplest way (at least for a physicist) to obtain the fermionic
representation is through the vertex operator construction.  It is
easy to see that this is the Coulomb gas partition function describing
a chiral bosonic field with exponential interaction. Introduce the
bosonic field $\varphi(z) $ with mode expansion
\eqn\henno{
\varphi(z) = \hat q+ \hat p\log  z +\sum_{n\ne 0} {H_{n}\over n} 
z^{-n}, }
\eqn\frt{
[ H_n, H_m]
=n\delta_{m+n, 0}; \ \ [\hat p,\hat q] = 1. }
and the vacuum state $|l\rangle$ defined by \eqn\ljai{
H_n|l\rangle =0 , \ \ (n>0); \ \ \ \hat p |l\rangle=l|l\rangle . }
The associated normal ordering is defined by putting $H_n, n>0$ to the
right.
Now define the vertex operators
\eqn\vop{V_{q}(z) = : e^{\varphi( q^{1/2} z)} : :e^{- \varphi( q^{-1/2} z)}: 
} satisfying the OPE
\eqn\opevo{
V_{q}(z) V_{q}(z') =
{(z-z')^{2} \over
(q^{1/2}z - q^{-1/2} z')( q^{-1/2}z - q^{1/2} z')} : V_{q}(z) 
V_{q}(z'):\ .}
Then the $\tau$-function \tauf\ can be written as the expectation
value
\eqn\ttatra{
\tau_{l} [t] = \Big\langle l\ \Big|\exp\left( \sum_{n> 0}t_nH_n 
\right)
\ {\bf g}\
\exp\left(-\sum_{n< 0}t_nH_n \right)\Big|\ l\Big\rangle 
}
where the operator ${\bf g}$ is defined as \eqn\ggege{{\bf g} =
\exp\left( e^{2\pi R\mu} \oint {dz\over 2\pi }	V_{q} (z) \right)
.}

\subsec{Fermionic representation }

The fermionic representation of the $\tau$-function now follows from 
the bosonization formulas \eqn\bBo{\psi(z)= :e^{-\varphi(z)}: , \ \ \ 
\psi^*(z)=
:e^{\varphi(z)}: , \ \ \
\partial\varphi(z)= :\psi^*(z)
\psi(z):
}
where 
\eqn\pzpo{
\psi(z) = \sum_{r\in \Z+ {1\over 2}}\psi_{r} z^{-r- {1\over 2}} , 
\quad
\psi^*(z) = \sum_{r\in \Z+ {1\over 2}}\psi^*_{-r} z^{-r- {1\over 2}}	}
is a chiral complex Ramond fermion field, whose modes satisfy the 
canonical anticommutation relations \eqn\cpmto{
[\psi_{r},\psi^*_{s}]_+=\delta_{rs}.
}

The Hamiltonians $H_n$ and the vertex operator \vop\ are represented 
as fermion bilinears
\eqn\curro{
H_{n} =
\sum_{r\in \Z+ {1\over 2}} :\psi^*_{r-n}\psi_{r}: \ \ (n\in \Z) }
\eqn\oomm{ V_{q}(z)= \psi(q^{1/2}z )
\psi^*( q^{-1/2} z ) .}
and the
vacuum states with given electric charge $l$ satisfy 
\eqn\mnfio{\eqalign{
\langle l|	\psi_{-r} &=\langle l| \psi^*_{r}= 0\ \ \ \
\ (r>l )\cr
\psi_{r}| l \rangle &=\psi^*_{-r}|l\rangle = 0\ \ \ \ 
\ (r> l). \cr}
}

\subsec{Hirota bilinear equations}

The ensemble of $\tau$-functions with different charge satisfy a set 
of bilinear equations \Hir .
Below we give a sketch of their derivation. 

The operator ${\bf g}$ defined by \ggege\ and \oomm\ is an exponential
of a fermionic bilinear and as such can be thought of as a
$Gl(\infty)$ rotation. The corresponding Lie algebra $gl(\infty)$ is
generated by all bilinears $\sum_{rs} a_{rs} \psi_{r} \psi^{*} _{s} $.
The operator
$$ S_{12} = \sum_{r\in\Z+{1\over 2}} \psi_{r} \otimes \psi^{*}_{r}
=\oint {dz\over 2\pi i} \psi(z) \otimes \psi^{*}(z) $$ is a fermionic
analog of the tensor Casimir in the sense that it commutes with the
tensor product of two copies of the operator ${\bf g}$:
\eqn\scom{ S_{12}\ {\bf g}\otimes {\bf g} = {\bf g}\otimes {\bf g}\
S_{12}.}  Further, the definition of the vacuum implies that $\psi_{r}
|\ l
\rangle \otimes \psi_{r}^{*} |\ l' \rangle =0$ ($l>l'$), since either $\psi_{r} 
|\ l \rangle=0$ or $\psi_{r}^{*} |\ l'\rangle =0$. Therefore
$$ S_{12} |\ l \rangle\otimes |\ l' \rangle=0$$ and similarly for the 
left vacuum.

Now we multiply eq. \scom\ from the left by $ \langle l +1\ | e^{ 
\sum_{n> 0}t_nH_n }
\otimes \langle l'-1\ | e^{ \sum_{n> 0}t'_nH_n } $ and from the right by
$ e^{ -\sum_{n< 0}t_nH_n } |\ l \rangle \otimes 
e^{ -\sum_{n< 0}t_nH'_n }|\ l' \rangle$
and commute the fernion operators with the exponents until they hit 
the left (right) vacuum and disappear.
As a result we obtain the following identities known as Hirota 
bilinear equations \Hir
\eqn\Hiilr{\eqalign{
&\oint_{C_\infty}{dz\over 2\pi i} z^{l-l'}
\exp\Big(\sum_{n>0} (t_n-t'_n) z^n\Big)
\tau_{l}(t-\tilde\zeta_+)
\ \tau_{l'}(t' +\tilde\zeta_+ ) =\cr
& \oint_{C_0}{dz\over 2\pi i} z^{l-l'}
\exp\Big(\sum_{n<0} (t_n-t'_n) z^{-n}\Big) \tau_{l+1}(t - \tilde\zeta_-)
\ \tau_{l'-1}(t' +\tilde\zeta_-) .
\cr}}
where
$$\tilde \zeta_+ = (\ldots, 0, 0, z^{-1}, z^{-2}/2, z^{-3}/3 , \ldots),
\ \ \tilde \zeta_- = (\ldots, z^{3}/3, z^{2}/2, z , 0, 0  \ldots).
$$
Expanding in $y_{n} = t'_{n} - t_{n} $ we obtain an infinite 
hierarchy of soliton partial differential equations. 
 \eqn\hireqn{\eqalign{&\sum_{j=0}^{\infty}P_{j+m}(-2y_+)P_j(\tilde 
 D_+)\exp
 \Big(\sum _{j\ne 0}y_jD_j\Big) \tau_{l+m+1}[t]\cdot \tau_l[t]\cr 
 &=\sum_{j=0}^{\infty}P_{j-m}(-2y_-)P_j(\tilde D_-)\exp \Big(\sum 
 _{j\ne 0}y_jD_j\Big) \tau_{l+m}[t]\cdot \tau_{l+1}[t]\cr}}
 where
 $$t_{\pm}=(t_{\pm 1},t_{\pm 2}, ...), \ y_{\pm}=(y_{\pm 1},y_{\pm
 2}, ...), $$
 $$ \tilde D_{\pm}= ( D_{\pm 1}, D_{\pm 2}/2, D_{\pm 3}/3, ...)$$
represent the Hirots's bilinear operators 
\eqn\rdr{D_n f[t]\cdot g[t]=
{\p\over \p x}f(t_n+x)g(t_n-x)\Big|_{x=0}, } 
and finally $P_{j} $ are
the Schur polynomials defined by 
\eqn\ssur{\sum_{j=1}^{\infty} P_j [
t]\ \lambda^j= \exp\Big(\sum_{n=1}^{\infty}t_n\lambda^n\Big).}  
The lowest, Toda equation, which is what we will actually need, is
obtained by equating to zero the coefficient in front of $y_{-1}$:
\eqn\TTtt{\tau_l\p_1\p_{-1}\tau_l-\p_1\tau_l\p_{-1}\tau_l+ 
\tau_{l+1}\tau_{l-1}
=0.}

\appendix{D}{Calculation of partition functions on  torus and
sphere }

\subsec{Solution of the equation 
for the sphere
}

Changing the  variable $y=e^u$ and  the function $X_0=-\phi-2u$ we
bring the equation \SEQ\ to the form
$$
\a \p_u^2\phi=(\p_u^2\phi+(\p_u\phi)^2+3\p_u\phi+2)e^\phi
$$
where it does not contain explicitely $u$.

By denoting $\psi(\phi)=\p_u\phi$
we obtain the first order ODE 
$$
(\a e^{-\phi}-1)\psi'_\phi\psi=(\psi+1)(\psi+2)
$$
with the solution
$$
{(\p_u\phi+1)\over(\p_u\phi+2)^2(e^{\phi}-\a)}=-C
$$
where C is the integration constant.
Solving the quadratic equation for $\p_u\phi$ and integrating it we
obtain:
$$
\phi+2u=
\int  d\phi{ 1 \over \sqrt{1-4C\a-4C e^\phi } }=
{2\over \sqrt{1-4C\a}}{\rm arcsinh}
\left(\sqrt{1-4C\a\over 4C} e^{-\phi/2}\right) + C_3 .
$$
Going back to original variable $y$ we see that fixing the couplings
$C=\left({2-R\over R}\right)^2$ and $C_3=0$ we reproduce \SFSOL\
generating the series in Moore's formula \ppbbmm .

\subsec{Solution of  the equation  for the torus}

The easiest is to change the variable from $y$ to
 $t=e^{{2-R\over R} X_0}$, according to the spherical solution \SFSOL .
 After a tedious but direct calculation 
\FIRST\ takes the form
$$
[t-(R-1)]\p_t f_1'+{R-1\over R-2}(1-1/t)f_1'=
{R(R-2)\over 12}t^{1/(2-R)} {(R-1)^2t-1\over [(R-1)t-1]^3}.
$$
It is a frst order inhomogenious ODE with the explicite solution
$$
f_1'= {R(R-2)\over 12}\int^t ds\ e^{B(s)-B(t)}\ s^{1/(2-R)}
{(R-1)^2s-1\over [(R-1)s-1]^3[s-(R-1)]}
$$
with
$$
B(t)={R-1\over R-2}\int^t dt {t-1\over t[t-(R-1)]}=
\log \left(t^{1/(R-2)}[t-(R-1)]\right).
$$
Fortunately (and may be due to the itegrability of the original Toda
equation) the resulting integral for $f_1$ can be explicitly
calculated:
$$
f_1'= {R(R-2)\over 12} {t^{1/(2-R)}\over t-(R-1)}\left({2-R\over
2(R-1)}{1\over[(R-1)t-1]^2}-{1\over (R-1)t-1} +D \right)
$$
where $D$ is an integration constant.

Using $f_1(y)=\int dy f_1'$ and integrating once more we get for the
partition function
\eqn\ZFIRST{\eqalign{ &f_1(y)=
-{1\over 24}\left({2D R\over R-1}+{R^2\over (R-1)^2}\right)\log  t \cr
& -{1\over 24}\left({2DR^2(R-2)\over R-1}-{2R^2-2R+1\over
(R-1)^2}\right)\log [ (R-1)-t] - {1\over 24}\log [1-t(R-1)] +C\log \l } }
where the last term is the zero mode of Toda equation with yet unknown
coefficient. 

To fix $D$ and $C$ let us use the boundary condition \FrenO\ for
$\l=0$
$$
f_1(\mu,\l)=- {R+R^{-1}\over 24} \log \mu,  \ \ \ \l=0. 
$$   
(Note that we always drop a constant, cutoff dependent, term from the
partition function.) Since $y\to \infty$ we have $t=e^{(1-2/R)\chi}\to
y^{R-2}$ and comparing the solution with the boudary conditions
\FrenO\ we conclude:
$$
2D{R(R-2)\over R-1}+{R^2(R-2)\over (R-1)^2}=R+R^{-1}.
$$

Note that, ``miraculously'', the second logarithmic term 
disappears from  \ZFIRST\ for this value of $D$. 

Since $F(\mu,\l$) is regular at $\l=0$  the constant $C$ 
should be chosen
as
$$
C=-{R+R^{-1}\over 12(2-R)}.
$$
With these values of $C$ and $D$  we finally obtain the result \GENZ .

  \listrefs

\bye